\documentclass[preprint,review,14pt]{cas-sc}

\usepackage[square,numbers]{natbib}
\usepackage{algorithm}
\usepackage{algpseudocode}
\usepackage{diagbox}
\usepackage{multirow}
\usepackage{float}
\usepackage{makecell}
\usepackage{graphicx} 
\usepackage{epsfig}
\usepackage{epstopdf} 

\def\tsc#1{\csdef{#1}{\textsc{\lowercase{#1}}\xspace}}
\tsc{WGM}
\tsc{QE}
\tsc{EP}
\tsc{PMS}
\tsc{BEC}
\tsc{DE}

\begin{document}
\let\WriteBookmarks\relax
\def\floatpagepagefraction{1}
\def\textpagefraction{.001}
\let\printorcid\relax

\shorttitle{A feature-preserving parallel particle generation method}

\shortauthors{}

\title [mode = title]{A feature-preserving parallel particle generation method for complex geometries}                      



%
\author[1]{Xingyue Yang}[style=chinese]

\fnmark[1]

\ead{yxy_nwpu@mail.nwpu.edu.cn}



\affiliation[1]{organization={School of Software, Northwestern Polytechnical University},
    city={Xi’an,  Shanxi},
    postcode={710129}, 
    country={China}}

\author[1,2]{Zhenxiang Nie}[style=chinese]
\ead{niezhenxiang@nwpu.edu.cn}
\fnmark[1]

\affiliation[2]{organization={Yangtze River Delta Research Institute of NPU Northwestern Polytechnical University},
    city={Taicang, Jiangsu},
    postcode={215400}, 
    country={China}}

\author[1,2]{Yuxin Dai}[style=chinese]
\ead{yuxin_dai@mail.nwpu.edu.cn}


\affiliation[3]{organization={Shenzhen Research Institute, Northwestern Polytechnical University},
    city={Shenzhen},
    postcode={518063}, 
    country={China}}

\author%
[1,2,3]{Zhe Ji}
\cormark[1]
\ead{jizhe@nwpu.edu.cn}

\cortext[cor1]{Corresponding author}

\fntext[fn1]{Xingyue Yang and Zhenxiang Nie contributed equally to this work.}

\setlength{\skip\footins}{5pt}



\begin{abstract}
\noindent In this paper, a Feature-preserving Particle Generation (FPPG) method for arbitrary complex geometry is proposed. Instead of basing on implicit geometries, such as level-set, FPPG employs an explicit geometric representation for the parallel and automatic generation of high-quality surface and volume particles, which enables the full preservation of geometric features, such as sharp edges, singularities and etc. Several new algorithms are proposed in this paper to achieve the aforementioned objectives. First, a particle mapping and feature line extraction algorithm is proposed to ensure the adequate representation of arbitrary complex geometry. An improved and efficient data structure is developed too to maximize the parallel efficiency and to optimize the memory footprint. Second, the physics-based particle relaxation procedure is tailored for the explicit geometric representation to achieve a uniform particle distribution. Third, in order to handle large-scale industrial models, the proposed FPPG method is entirely parallelized on shared memory systems and Boolean operations are allowed to tackle structures with multiple assemblies. Intensive numerical tests are carried out to demonstrate the capabilities of FPPG. The scalability tests show that a speedup of ~10X is achieved through multi-threading parallelization with various models. Comparative studies with other particle generation methods show that FPPG achieves better performance in both runtime and accuracy. Last, two industrial cases of vehicle wading and gearbox oiling are studied to illustrate that FPPG is applicable to complex geometries. 

\end{abstract}


\begin{highlights}
\item Explicit geometric representation combined with customized data structures and algorithms fully preserves geometric features and ensures particles are uniform and body-fitted.
\item Parallelization on shared memory hardware with OpenMP achieves a speedup of over 10X, enabling rapid particle generation and optimization.
\item Numerous validation and comparative numerical experiments comprehensively verify the feasibility, parallelism, accuracy and robustness of the algorithm.
\item Vehicle wading and gearbox splash lubrication case demonstrate that the particle configurations generated by the algorithm provide more reliable performance in industrial simulations.
\end{highlights}


\maketitle

\section{Introduction}
Numerical simulation is a powerful tool for analyzing and predicting complex physics problems. Mesh-based numerical methods rely on meshes to represent the entire system, such as the Finite Difference Method (FDM)\cite{perrone1975general} and the Finite Element Method (FEM)\cite{idelsohn2004particle}. Despite advances in mesh generation techniques and computer hardware, creating high-quality meshes remains time-consuming. This limits the accuracy, robustness and efficiency of simulations\cite{ito2013challenges, chawner2016geometry, zhang2022smoothed}. Additionally, in problems involving free surfaces, deformable boundaries, and multi-scale phenomena, connectivity between mesh elements becomes a significant challenge, resulting in numerical errors\cite{chen2017meshfree, bishop2020kinematic}. In contrast, meshfree methods represent the computational domain and boundaries with particles, thus avoiding the inherent limitations of mesh-based methods. With the improvement of meshfree methods, various approaches have been developed. Point cloud-based meshfree methods such as the Element Free Galerkin (EFG)\cite{belytschko1994element, li2023theoretical} and the Radial Basis Function (RBF)\cite{liu2023adaptive, hosseini2021meshless}, and particle-based meshfree methods, such as Smooth Particle Hydrodynamics (SPH)\cite{shrey2024smooth, bagheri2024review}  and Particle Finite Element Method (PFEM)\cite{rizzieri2024simulation, cremonesi2020state}, have demonstrated significant advantages over mesh-based methods in certain applications\cite{sandin2024particle, zhang2024state, patel2020meshless, bui2021smoothed}. Particle-based method based on a Lagrangian framework in which particles adaptively follow the movement of the material. This capability is particularly effective when handling large deformations\cite{navas2021explicit, nguyen2021efficient}, free surfaces\cite{wang2023compact, yan2024updated}, multiphase flows\cite{peng2021particle, yan2020higher}, and complex boundaries\cite{matsunaga2020improved, han2024algorithm}. Furthermore, the pairwise particle computations enable efficient parallelization on multi-core and distributed computing systems\cite{vacondio2021grand, zhu2023novel}. As a result, meshfree methods have found widespread application in various scientific and engineering fields, including computational astrophysics\cite{alonso2023mesh, schaller2024swift}, computational fluid dynamics\cite{mazhar2021meshfree, belytschko2023meshfree}, solid mechanics\cite{al2024implementation, littlewood2024peridigm}, biomedicine\cite{mountris2023cardiac, barbosa2022computational, duckworth2021smoothed}, and materials science\cite{liu2022elastoplastic, liu2023meshfree}.

Particles play an indispensable role in meshfree methods\cite{xiao2022development, zhang2023initial}. Nevertheless, most existing particle-based methods ignore or do not prioritize achieving high-quality particle distribution in the initial stage. To attain the desired distribution, constraints are placed throughout the evolutionary process\cite{suchde2021meshfree, zhao2023role, khayyer2019projection}. However, both the computational accuracy and the convergence rate are compromised if the initial conditions do not ensure uniform particle distribution and cannot precisely define the boundaries of the computational domain\cite{fu2019optimal, zhu2021cad}.

Existing initial particle generation methods can be divided into two categories. The first category uses the vertices or centroids of unstructured meshes. For instance, the open-source software RKPM2D\cite{huang2020rkpm2d} generates initial points through Voronoi diagram partitioning. This method imposes fewer quality requirements on the mesh and does not require any subsequent post-processing or optimization of the particle distributions. However, for complex geometries, mesh generation may not work, which would result in low robustness. The second category uses regular lattices that generate particles at points such as the body-centered or face-centered points of a cubic lattice structure. For instance, DualSPHysics' particle generator Gencase\cite{dominguez2011development} generates particles on the lattice to ensure uniform distribution. However, the points near the boundaries are highly irregular and therefore complex geometries require high resolution. To better fit geometries, particles can be generated at the centers of tetrahedral or hexahedral volume elements\cite{meng2021modeling, liu2023modeling}, allowing accurate geometric representation, but the particle distribution remains non-uniform.

Various improvement methods have been proposed to generate a high-quality initial particle distribution that is uniform and body-fitted. Diehl et al.\cite{diehl2015generating} developed the Weighted Voronoi Tessellation (WVT) algorithm, which redistributes particles based on proximity to neighbors and desired resolution to achieve uniform distribution. Fu et al.\cite{fu2019optimal} optimized an energy function and applied the level-set method to describe geometries, using ghost particles to manage boundary conditions. Ji et al.\cite{ji2021feature} introduced a feature definition system with a boundary correction term to address incomplete kernel support near surfaces, enabling a more accurate representation of complex geometric features. Zhu et al.\cite{zhu2021cad} constructed an implicit zero level-set function and used a surface bounding method to constrain surface particles, ensuring that the distribution accurately reflects the geometry and maintains uniformity. Yu et al.\cite{yu2023level} focused on preprocessing for complex geometries, using level-set-based re-distance algorithms and cleaning up “dirty” geometries to improve particle distribution around sharp features. Notably, these methods are primarily based on implicit geometric frameworks, which require constructing a signed global distance function and incur significant computation costs. Accurately capturing geometric features such as sharp corners, ridgelines, and singularities remains challenging and inevitably leads to a loss of modeling accuracy, especially for complex geometries. Few methods are able to achieve high-quality initial particle distributions.

Explicit geometric representation addresses the aforementioned issues by defining the geometry directly based on the input, eliminating the need to construct a global distance function. In addition, implicit methods require a conversion process based on the input, which inevitably results in a loss of precision. For instance, level-set methods approximate shapes using signed distance functions and typically yield smooth boundaries. This can blur sharp edges or fine features, causing important geometric details to be lost. In contrast, explicit method generates particles directly on geometric surfaces, avoiding transformations and preserving accuracy, which is essential for complex models.

Although explicit geometric representation can accurately describe features, achieving uniform particle distribution remains a challenge. Due to the meshfree nature, applying precise and robust boundary conditions is difficult. Generating ghost particles\cite{fu2020adaptive} at boundaries prevents particle penetration, but requires additional particles to maintain mesh quality, increasing computational cost and memory usage. Ji et al.\cite{ji2021feature} propose a feature definition system that implicitly improves mesh quality through pairwise particle forces without requiring additional constraints to prevent particle penetration. However, the process is not fully automated and a special preprocess is required to resolve the mismatch between the level-set function of the background mesh discretization and the feature lines. 

This paper proposes a particle generation method based on explicit geometric representation to achieve parallel and automatic generation of high-quality particles. First, an efficient and improved data structure is developed that directly uses geometric information to avoid loss of precision due to transformations. Volume partitioning and subdivision are used to define and classify different features for management and storage. Second, a particle mapping and feature line extraction algorithm is proposed. Surface particles are mapped to achieve body-fitted distribution and feature line particles are generated to capture geometric details, ensuring a comprehensive representation of arbitrary complex geometry. Next, a physics-based particle relaxation method tailored for explicit geometry is introduced. Based on the feature system, different forces are applied to achieve uniform distribution. A mapping algorithm is employed to maintain body-fitted characteristics in surface particle relaxation. In addition, a narrow-band technology is introduced to optimize volume particles close to the geometry boundary, thereby reducing the computational consumption. Finally, the proposed FPPG method is implemented with the OpenMP parallelization technique to achieve both goals of high-quality and rapid particle generation.

The paper is organized as follows: Section 2 provides a brief overview of the feature-aware SPH method presented in \cite{ji2021feature}. Section 3 provides a detailed introduction to the proposed feature-preserving particle generation method, describing explicit geometric representation, fully feature-preserving, body-fitted particle generation, and uniform distribution of particle optimization. In Section 4, numerical examples demonstrate that FPPG achieves high-quality particles for complex geometries. Additionally, performance tests confirm the remarkable parallelism and scalability of FPPG. Two industrial vehicle wading and gearbox oiling cases are examined to illustrate that FPPG is applicable to complex geometries.

\section{The feature-aware SPH method}

Before introducing the proposed FPPG method, the feature-aware SPH method for particle generation is first briefly overviewed \cite{ji2021feature} in this section.

\subsection{Geometry definition}

The feature-aware SPH method represents geometry through a zero level-set function \cite{osher1988fronts}
\begin{equation}
     \varGamma = \{ (x,y)|\phi (x,y,t) = 0\}. \label{eq1}
\end{equation}

The volume mesh generation region is defined as the positive phase, i.e. $\varGamma_{+} = \{ (x,y)|\phi (x,y,t) > 0\}$. The surface mesh generation region corresponds to the zero level-set, i.e. $\varGamma_{0} = \{ (x,y)|\phi (x,y,t) = 0\}$. Moreover, additional inputs are used to define sharp edges and singularities.

\subsection{Feature definition}

To characterize the features, five categories of feature cells are defined on a Cartesian background mesh, i.e. positive cell, negative cell, feature-surface cell, feature-curve cell, and singularity cell. The particle categories are determined by the cells they occupy. They are categorized into four types, i.e. singular particles $(\mathbb{P}_{si})$, feature-curve particles $(\mathbb{P}_{C})$, feature-surface particles $(\mathbb{P}_{s})$, and positive particles $(\mathbb{P}_{+})$, corresponding to singular points, sharp edges, geometric surfaces, and interior particles, respectively.

\begin{figure}[pos=H]
	\centering
		\includegraphics[scale=.8]{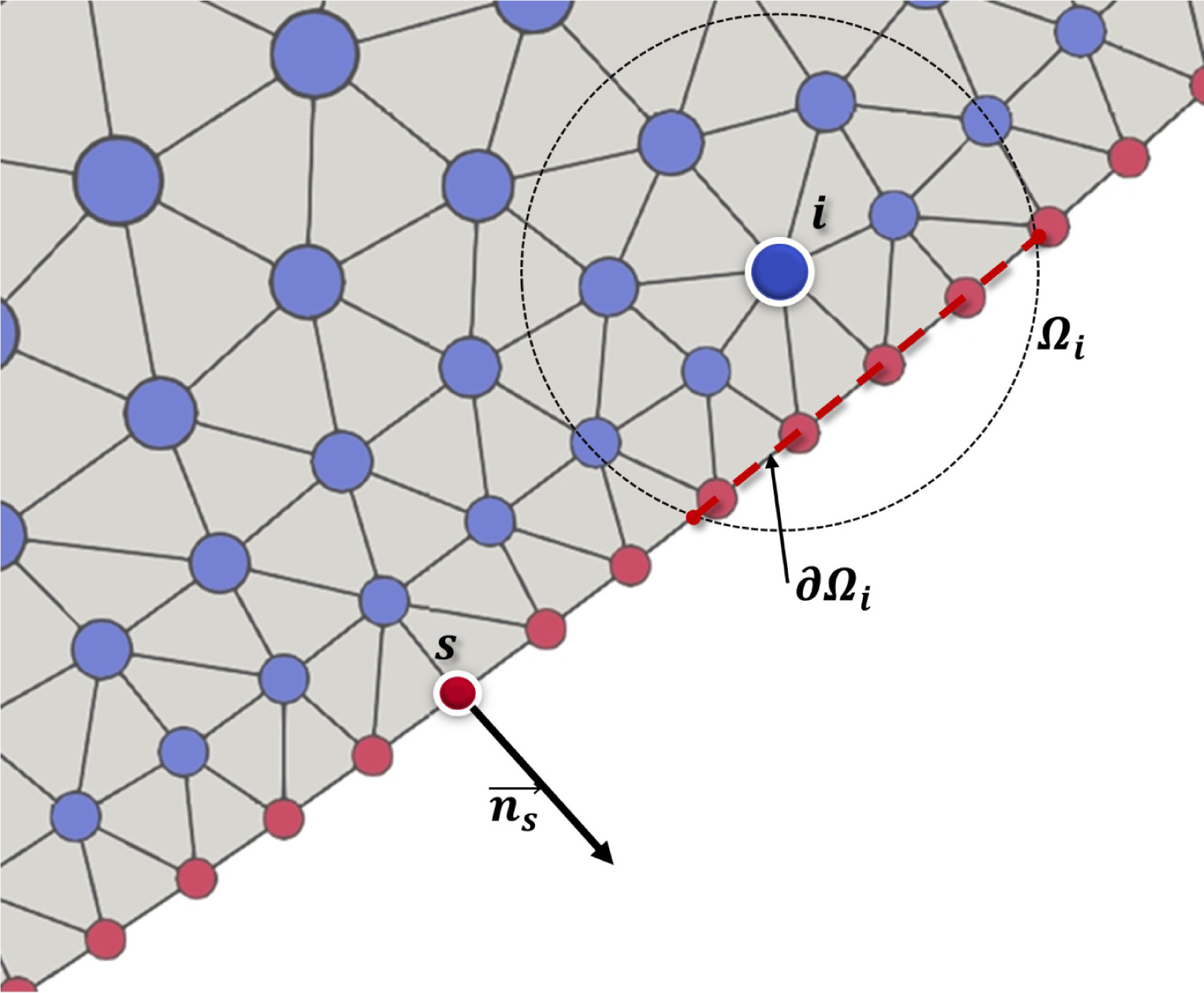}
	\caption{Interaction with boundary particles and illustration of inconsistency in the kernel support region. (From \cite{ji2021feature}).}
	\label{fig1}
\end{figure}

Based on the tag system defined above, interactions between different feature particles can be set. For example, $\mathbb{P}_{C}$ exerts a repulsive force on $\mathbb{P}_{s}$ and $\mathbb{P}_{+}$ to prevent penetration, while $\mathbb{P}_{si}$ is the boundary condition for all other particle types. Figure \ref{fig1} illustrates the interaction between $\mathbb{P}_{s}$ (red points) and $\mathbb{P}_{+}$ (blue points) in a two-dimensional scenario. Within the cutoff range (dashed circle) of particle i (highlighted blue point), all red particles are considered boundary conditions, which act as a non-penetration constraint and compensate for the insufficient core support near the boundary.

\subsection{Modeling equations and numerical discretization}

The evolution of the mesh vertices is calculated based on the fluid relaxation process and modeled as an isothermal compressible flow. The Lagrangian form of the governing equations is
\begin{equation}
    \frac{d\rho }{dt} = -\rho  \bigtriangledown \cdot \mathbf{v}, \label{eq2}
\end{equation}
\begin{equation}
    \frac{d\mathbf{v} }{dt} = -\mathbf{F}_p + \mathbf{F}_v ,\label{eq3}
\end{equation}
\begin{equation}
   \frac{d\mathbf{x} }{dt} = \mathbf{v} ,\label{eq4}
\end{equation}
where $\rho$ is the density, $\mathbf{v}$ is the velocity vector and $\mathbf{x}$ is the position. $\mathbf{F}_p$ and $\mathbf{F}_v$ denote the pressure and viscous forces, respectively. To close the system, EOS is required

\begin{equation}
   p = P_0(\frac{\rho }{\rho _t})^\gamma ,\label{eq5}
\end{equation}
where $P_0$ is a constant pressure and $\gamma > 0$ is a user-defined parameter. This EOS drives the particles to relax to the target distribution.

The model equations are discretized and solved using the Smooth Particle Hydrodynamics method. Assuming $\gamma = 2$ in the equation of state, the momentum equation is discretized as

\begin{equation}
   \frac{d\mathbf{v}}{dt} = -\displaystyle\sum_{j}m_j(\frac{p_0}{p_{t,i}^2}+\frac{p_0}{p_{t,j}^2})\frac{\partial W(r_{ij},h_{ij})}{\partial r_{ij}} \mathbf{e}_{ij}+\displaystyle\sum_{j}m_j\frac{2\eta _i\eta _j}{\eta _i + \eta _j}(\frac{1}{p_{t,i}^2}+\frac{1}{p_{t,j}^2})\frac{\partial W(r_{ij},h_{ij})}{\partial r_{ij}}\frac{\mathbf{v}_{ij}}{r_{ij}} ,\label{eq6}
\end{equation}
where h is the smoothing length of the kernel function, $W(r_{ij},h_i)$ is the kernel function, $\frac{\partial W(r_{ij},h_{ij})}{\partial r_{ij}}$ is derivative of kernel, $\mathbf{r}_{ij} = \mathbf{r}_i - \mathbf{r}_j$ is the connection vector between particle i and j, $\mathbf{e}_{ij} = \frac{\mathbf{r}_{ij}}{r_{ij}}$is the unit vector of $\mathbf{r}_{ij}$, $\mathbf{v}_{ij} = \mathbf{v}_i - \mathbf{v}_j$, $h_{ij} = \frac{h_i+h_j}{2}$ is the average smoothing length of particle i and j, $\eta = \rho \nu $ the dynamic viscosity, $\nu \sim 0.1r_c|\mathbf{v}|$, where $r_c$ is the cut-off radius of particle interaction range.

For particles without kernel support near the boundary, the reformation term $\gamma (x)$ is introduced. The $\gamma (x)$ of particle i can be calculated directly from the target volume of particle j, similar to the discrete Shepard coefficients \cite{shepard1968two}.

\begin{equation}
    \gamma _i = \displaystyle\sum_{j}W(\mathbf{r}_{ij},h_i)V_{t,j}, \label{eq7}
\end{equation}
where the summation is carried out over all neighboring particles of i.

Finally, the pressure force term can be rewritten as

\begin{equation}
    \mathbf{F}_{p.i} = \frac{1}{\gamma _i}\displaystyle\sum_{j}m_j(\frac{p_0}{p_{t,i}^2}+\frac{p_0}{p_{t,j}^2})\frac{\partial W(r_{ij},h_{ij})}{\partial r_{ij}}\mathbf{e}_ij+\frac{1}{\gamma _i}\displaystyle\sum_{b}(\frac{p_0}{p_{t,i}^2}+\frac{p_0}{p_{t,b}^2})W(\mathbf{r}_{ib},h_{ib})\mathbf{n}_bA_{t,b} , \label{eq8}
\end{equation}
realization details can be consulted in paper \cite{ji2021feature}.

From a physical perspective, the additional term introduced in the pressure can be interpreted as a "repulsive force" of boundary particles, which prevents particles of the positive phase from penetrating into the negative phase. This significantly increases the stability and robustness of the system. In the momentum equation, the additional term dynamically adjusts the positions of particles near the boundary, continuously enforcing the non-interpenetration boundary condition.

Defining a feature system prevents particles from penetrating domain boundaries without additional constraints and implicitly improves mesh quality through pairwise particle forces. An additional advantage of this method is that it is fully parallelizable. However, this method is based on a zero-level-set function, which can introduce considerable error when discretizing complex geometries and requires specific input information. Furthermore, constructing a global signed distance function results in high memory consumption. Inspired by the feature system, a new method based on explicit geometric representation, FPPG, is proposed and described in detail in Section 3.

\section{The feature-preserving particle generation method}

\subsection{Explicit geometric representation}

Unlike particle generation methods that rely on implicit geometric representations, the Feature-Preserving Particle Generation (FPPG) method employs an explicit geometric representation. STL files that describe the geometry in terms of triangular facets are often used as input. The geometric volume is constructed on a Cartesian background grid, with subdivision reducing complexity. Following \cite{ji2021feature}, different grid categories are defined based on the spatial relationship between the background grid and the triangular facets. An efficient data structure is designed to optimize storage efficiency and reduce the high storage requirements associated with complex geometric data.

\subsubsection{Geometric volume partitioning and subdividing} \label{geometric volume partitioning and subdividing}

\begin{figure}[pos=H]
	\centering
		\includegraphics[scale=.48]{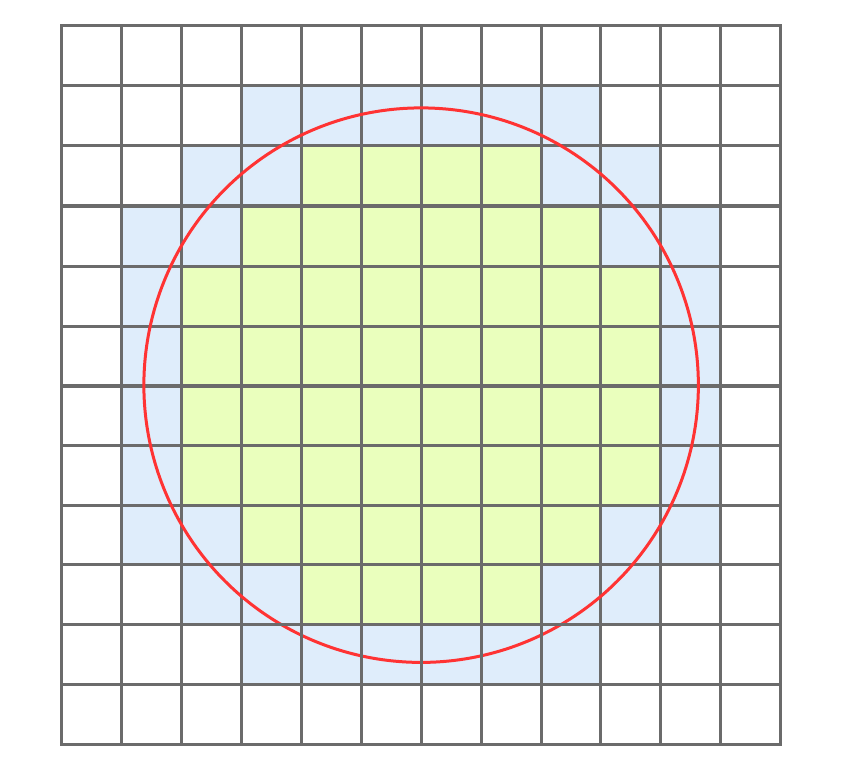}
	\caption{Geometric volume partition. Red is the geometric surface, blue is the surface mesh and green is the internal mesh.}
	\label{fign-2}
\end{figure}

Geometric volume partitioning divides the volume into non-overlapping sub-volumes to enable efficient organization and management of geometric data for rapid query and search. Various methods can be used for partitioning, such as uniform grids, octrees and KD trees. This work does not consider multi-resolution so the particle size is consistent. Therefore, a uniform grid partitioning based on the Cartesian coordinate system is employed. Figure \ref{fign-2} illustrates a simple 2D division.

To accurately represent complex geometric shapes, the background grid of the geometric volume needs to correspond to the complex geometric surface. The STL file approximates the surface of any 3D geometry with continuous triangles. Triangle facet information can be obtained directly from the STL file. Higher resolution requires a smaller background grid. However, for large spans of triangular facets, multiple meshes are involved, including those that do not intersect with the triangular facets, resulting in unnecessary computation time and memory overhead. To reduce the span of the triangles, a subdivision process is introduced.

\begin{figure}[pos=H]
\centering
\includegraphics[scale=.6]{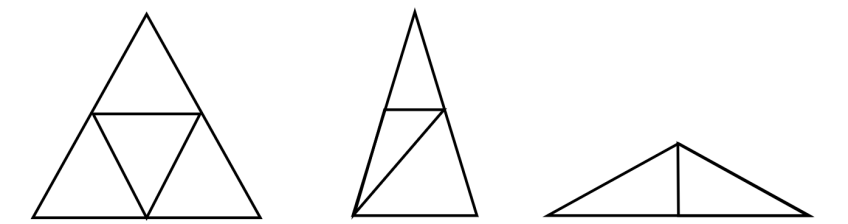}
\caption{Three basic classes of triangular decompositions.}\label{fign-3}
\end{figure}

The subdivision principle is straightforward: find the midpoints of the three edges of the triangle. New triangles are formed by connecting the vertices of the original triangle to the midpoints of its edges, thereby dividing the original triangle into four smaller triangles. However, due to the varying shapes and sizes of triangles, the subdivision is not always uniform and can be categorized into three types. As shown in Figure \ref{fign-3}, a triangle can produce 4, 3, or 2 new triangles after a single subdivision. To standardize the subdivision process, the concept of virtual points is introduced. If the length of the triangle edge already meets the resolution requirement, further subdivision is considered unnecessary and its midpoint is called a virtual point. Subtriangles are considered nonexistent if one of their vertices is virtual points.

\begin{figure}
\centering
\includegraphics[scale=.8]{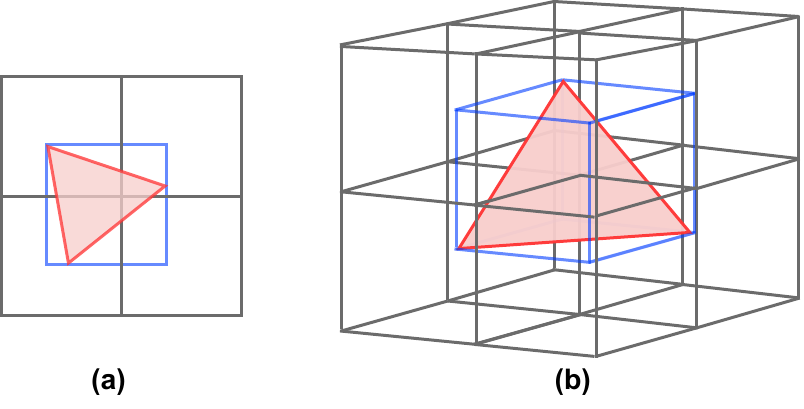}
\caption{The relationship between the size of the triangles and the background grid. (a) Two-dimensional. (b) Three-dimensional.}\label{fign-4}
\end{figure}

\begin{figure}
\centering
\includegraphics[scale=.55]{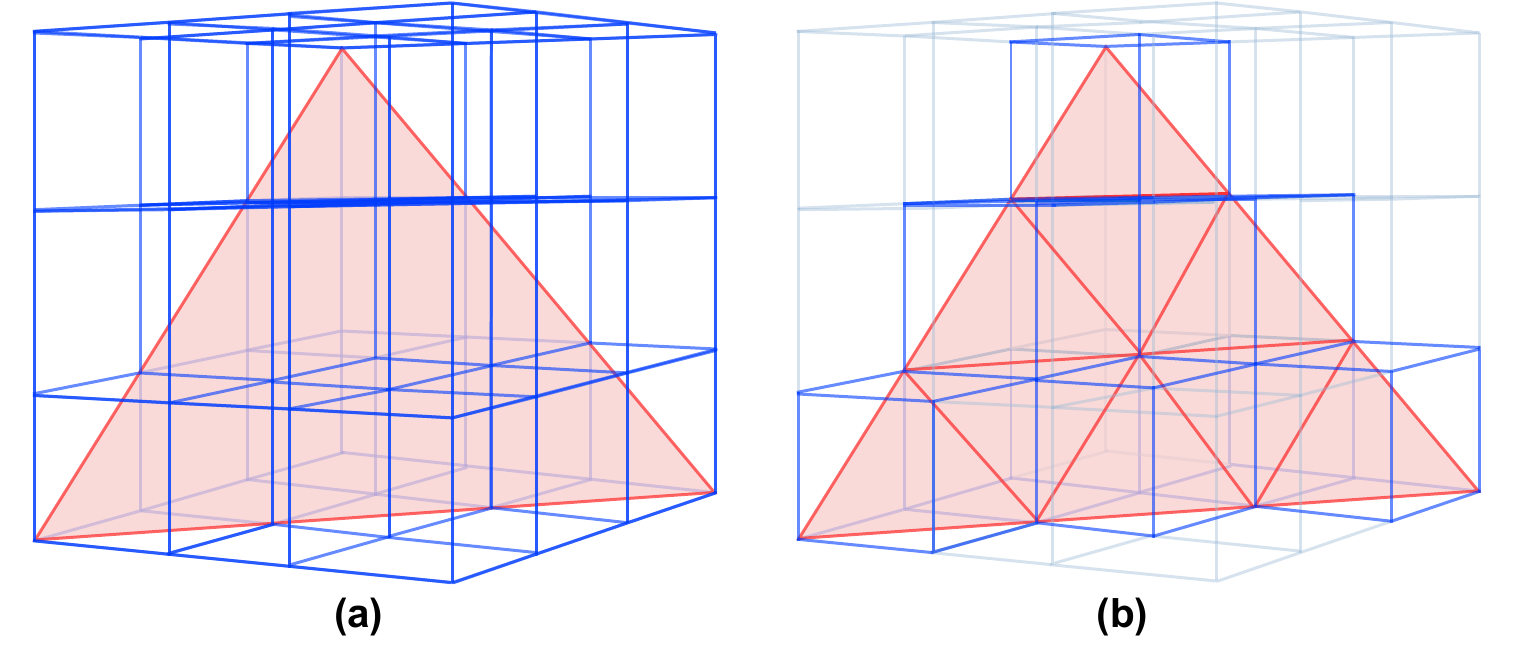}
\caption{Schematic representation of a grid of intersecting triangles. (a) Unsubdivided triangle. (b) Subdivided triangle.}\label{fign-5}
\end{figure}

In this paper, the size of the triangular facet is represented by its Axis-Aligned Bounding Box (AABB). Once divided, it is essential to ensure that the AABB length does not exceed the dimensions of the background grid. As shown in Figure \ref{fign-4}, if the AABB of a triangle in 2D does not exceed a single grid cell, it can intersect with up to four neighboring background grid cells. In 3D, the intersecting grid cells will not exceed eight. The subdivision algorithm is recursively applied to each triangle until all subtriangles meet the specified size requirements. Figure \ref{fign-5} illustrates the intersection between triangular facets and the background grid in the 3D volume. In Figure \ref{fign-5} a), the triangle is unrefined and its AABB spans 27 background grids. In Figure \ref{fign-5} b), the intersecting background grids are reduced to 9 after refinement. This effectively eliminates unnecessary grids and increases computing efficiency.

\subsubsection{Feature system} \label{feature system}

Following \cite{ji2021feature}, the background grid is classified into three categories: surface grid, internal grid and external grid. A fast 3D triangle-box overlap testing algorithm \cite{akenine2005fast} is introduced to efficiently determine whether triangle facets intersect with the background grids. Subdividing the triangle faces reduces the scale of the problem. Surface grids are identified by directly searching the grids that the AABB of smaller facets intersects.

\begin{figure}[pos=H]
\centering
\includegraphics[scale=1.2]{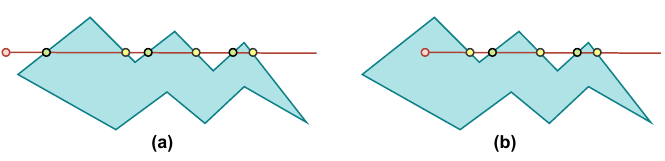}
\caption{Diagram of the raycasting algorithm. (a) The point lies outside the polygon and emits rays that cross the polygon an even number of times. (b) The point lies within the polygon and emits rays that cross the polygon an odd number of times. To simplify the calculation, a horizontal or vertical direction is usually chosen to emit the rays. }\label{fign-6}
\end{figure}

The raycasting algorithm is employed to define internal and external grids, as detailed in Figure \ref{fign-6}. In the 3D volume, the XY plane is treated as one dimension and the Z-axis as the other. A ray is thrown from the lowest point along the Z-axis in the positive direction. The grids in the XY plane are traversed sequentially along the Z-axis. A line segment is drawn from the center of each grid to the starting point. If the number of intersections between the line segment and the triangle faces is odd, the grid is an internal grid; if it is even, it is an external grid.

\subsubsection{Efficient data structure} \label{efficient data structure}

The primary limitation of explicit geometric representation is the substantial storage requirements. In the 3D volume, grid positions can be uniquely determined by the x, y and z coordinates, and grid information is usually stored in a 3D matrix. However, the computational domain of interest, consisting of surface and internal grids, generally represents only a small portion of the volume. Much of the grid information within the computational domain is redundant. Storing this information entirely would result in excessive and unnecessary memory consumption, especially for large geometric volumes. Therefore, the matrix used to store grid information is a sparse matrix.

To efficiently store sparse grid information, a Morton space-filling curve is employed to map high-dimensional spatial data into one-dimensional space. This approach reduces storage requirements while preserving the locality of the original data. More importantly, it provides an efficient indexing method that allows direct identification of a target grid cell by calculating the Morton code from the three-dimensional coordinates. The conversion between three-dimensional coordinates and the Morton code is given by Equation (\ref{eq9}).

\begin{equation}
    \bar{x} = \begin{pmatrix}
...\bar{x}_x^3 \bar{x}_x^2 \bar{x}_x^1 \bar{x}_x^0\\ 
...\bar{x}_y^3 \bar{x}_y^2 \bar{x}_y^1 \bar{x}_y^0\\ 
...\bar{x}_z^3 \bar{x}_z^2 \bar{x}_z^1 \bar{x}_z^0
\end{pmatrix}\rightarrow Z(\bar{x})=\cdots \bar{x}_z^3\bar{x}_y^3\bar{x}_x^3 \bar{x}_z^2\bar{x}_y^2\bar{x}_x^2 \bar{x}_z^1\bar{x}_y^1\bar{x}_x^1\bar{x}_z^0\bar{x}_y^0\bar{x}_x^0 . \label{eq9}
\end{equation}

After marking the surface grids, it is necessary to store the mapping between grids and corresponding triangular facets. A single grid may intersect multiple triangular facets, and searching for all facets corresponding to a grid requires traversing the entire mapping, which is inefficient. To enable fast retrieval, a sparse spatial data structure based on a compact hash function is constructed. First, using the locality of the Morton code, the mapping relations are sorted by the Morton code, and all triangles corresponding to the same grid are stored contiguously. Next, a compact list, $C_{compact}$, is constructed, where $C_{compact}^{begin}$ records the memory location of the first triangle in the grid, and $C_{compact}^{length}$ records the number of triangles in the grid. This compact list allows quick access to all triangles that correspond to a specific grid.

To reduce the complexity of random access to arbitrary elements, a hash function is introduced. The formula for calculating the hash value is 
\begin{equation}
    H(\bar{x}) = (p_1\bar{x}_x+p_2\bar{x}_y+p_3\bar{x}_z)\%H_{size} , \label{eq10}
\end{equation}
where $p_1=738560931$, $p_2=19349663$, $p_3=83492791$, $H_{size}$is the length of the hash table, $\bar{x}$ is the grid coordinate calculated from the particle coordinates.

\begin{figure}[pos=H]
\centering
\includegraphics[scale=.8]{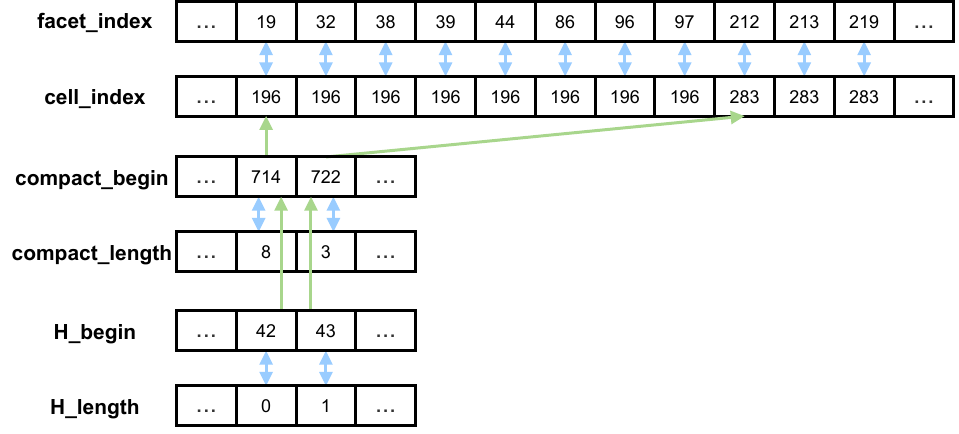}
\caption{The sparse spatial data storage structure based on the compact hash function. }\label{fign-7}
\end{figure}

To prevent potential "collisions" in hash mapping, radix sort is applied to reorder $C_{compact}$ and ensure that grids with the same hash value are stored in adjacent memory locations. Similarly, a compact list is constructed where $H_{begin}$ stores the index position in $C_{compact}$ of the first grid with the same hash value and $H_{length}$  stores the number of grids with the same hash value. The sparse spatial data storage structure based on the compact hash function is shown in Figure \ref{fign-7}. This structure allows rapid retrieval of all triangular facet information intersecting a grid based on the three-dimensional coordinates of any point in the volume.

\subsection{Body-fitted and feature-preserving particle generation} \label{Body-fitted and feature-preserving particle generation}

In this paper, boundary particles representing the surface of the geometry are termed surface particles, while particles within the enclosed volume are termed volume particles. Initially, particles are generated at the centers of the marked surface and internal grids. However, since the center of a surface grid typically does not lie exactly on the surface, mapping surface particles is necessary to maintain the body-fitted property. In contrast, internal grids do not require body-fitted, and the centers can be used directly. For complex geometries with sharp edges and singular points, simply using grid centers may result in losing geometric features. Consequently, feature extraction is used to add feature particles to maintain these features.

\subsubsection{Mapping and body-fitted particles} \label{mapping and body-fitted particles}

\begin{figure}[pos=H]
\centering
\includegraphics[scale=.8]{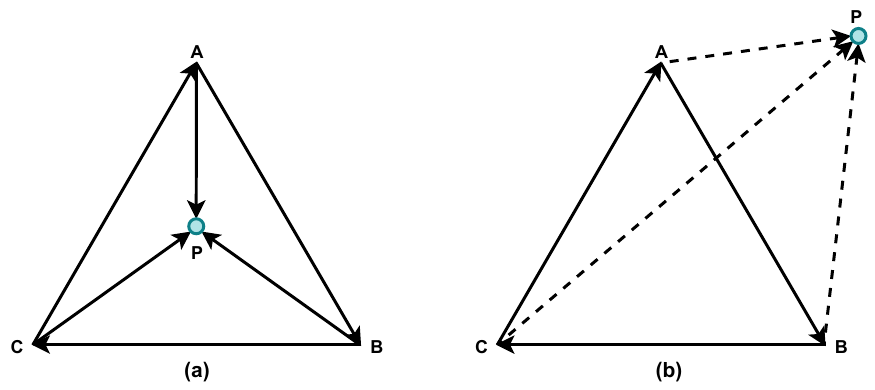}
\caption{Fork multiplication determines whether the point lies inside the triangle. The three vertices of a triangle are A, B and C. P is the point. Calculate the cross product of each side and each vertex to the point P. If the directions are all the same, (a) P lies inside the triangle; otherwise (b) P is not inside the triangle. }\label{fign-8}
\end{figure}

Particles generated at the centers of surface grids are unlikely to fit the geometric surface exactly and cannot be used directly. A mapping method ensures that the particles accurately fit the geometry. In this method, the grid center is projected to the nearest point on the surface, and the coordinates of the projection point are used as the surface particle coordinates. The geometric surface consists of triangular facets and the projection occurs along the normal vector of the triangle. The projection point must satisfy two criteria: first, the point must lie within the triangle; second, the distance between the grid center and the projection point must be minimized. Using the cross product method, the first condition can be checked directly, as illustrated in Figure \ref{fign-8}.

\begin{figure}[pos=H]
\centering
\includegraphics[scale=1]{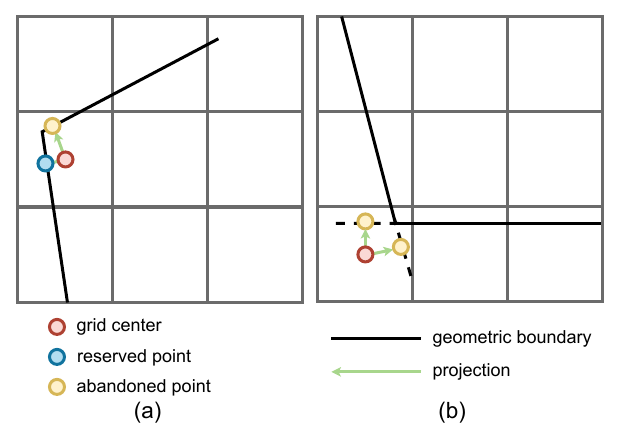}
\caption{Two-dimensional scheme of a point projection onto a geometric boundary. (a) The grid center is successfully projected and the projection point with the shortest distance is selected. (b) The projection fails and none of the projection points lie on the geometric boundary, so this projection is discarded. }\label{fign-9}
\end{figure}

The second condition is to use the efficient data storage structure to retrieve all triangle facets that intersect the grid. The distances from the grid center to each triangle are then calculated and sorted. The triangle with the shortest distance is selected for projection. If the projection point is outside the triangle, it is rejected and the process continues with the nearest triangle. If no projection point falls within a triangle, the grid is discarded. Figure \ref{fign-9} illustrates a simple example. In Figure \ref{fign-9} a), the grid center (red particle) can be projected onto both the yellow and blue locations, keeping the nearest blue location. In Figure \ref{fign-9} b), the grid center (red particle) cannot be projected onto any of the intersecting triangles and is therefore discarded.

\subsubsection{Feature extraction and feature line particles} \label{feature extraction and feature line particles}

\begin{figure}[pos=H]
\centering
\includegraphics[scale=1]{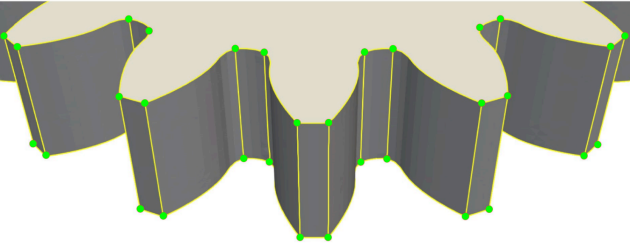}
\caption{Gear model sharp edges and singularities. }\label{fign-10}
\end{figure}

During the particle mapping phase, grid centers near sharp edges and singularities may not be accurately projected onto the surface, resulting in missing or inaccurate particles. As shown in Figure \ref{fign-9}(b), the red particle is discarded after projection, leading to the loss of geometric features, especially sharp corners within the grid. As a result, the existing particles cannot accurately represent the geometric features, especially complex geometries. Figure \ref{fign-10} shows a portion of a gear model, where the yellow lines represent feature edges and the green dots represent feature points. In these sharp regions  there is a risk that particles will be missed. To preserve the geometric properties, feature extraction is applied, in which feature particles are generated to complement the initial set of particles.

Feature extraction involves identifying feature edges and singular points based on complex geometric shapes and thus constructing feature lines. These feature lines represent geometric characteristics, such as edges, holes, protrusions, and fillets. Proper definition of feature lines can improve the quality and adaptability of the particle and ensure the precision of simulations and analysis. This study uses a dihedral angle-based method, which is well suited for rapid extraction of feature lines while preserving geometric details.

\begin{figure}[pos=H]
\centering
\includegraphics[scale=1]{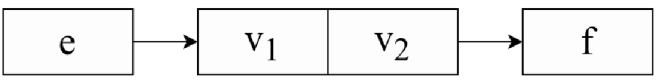}
\caption{Edge-to-face indexing relationships. }\label{fign-11}
\end{figure}

First, a mapping between edges and faces is established to facilitate the calculation of the dihedral angle and ensure storage uniqueness. In Figure \ref{fign-11}, $e$ represents the index of the edge, $v_1$ and $v_2$ are the indices of the two vertices of the same edge, where $v_1 < v_2$ to ensure uniqueness in storage, and $f$ represents the index of the triangle face. Since the order of edge indices is unimportant, the calculation of the map $map_1$ ( $e \rightarrow  v_1$ ) can be performed after the calculation of the map $map_2$ ( $v_2 \rightarrow f$ ) in the algorithm implementation. After obtaining $map_2$, $map_1$ only needs to be calculated sequentially, effectively reducing the complexity of the algorithm implementation.

Next, the feature edges and singular points are marked in the following steps. The normal vectors of all faces and the dihedral angles on the edges are calculated. A normal vector is obtained by computing the cross product of any two edge vectors of a facet. To ensure the normal direction from inside to outside, the angle between the normal vector and the viewpoint must be checked to determine the orientation of the normal. For each edge, the two connected faces are accessed (if an edge is connected to only one face, it is directly classified as a feature line). The normal vectors of these faces are then used to calculate the dihedral angle of the current edge. This angle is compared to a predefined threshold, where edges with dihedral angles below the threshold are excluded and marked as feature edges. Finally, all feature edges are traversed and the endpoints are counted. Endpoints with an access count of one or more than two are identified as singular points.

Finally, feature lines are constructed, followed by feature particle generation. The process begins at each singular point. Depth-first search (DFS) is utilized to identify the feature edges connected with that singular point. To prevent revisiting edges, a coloring technique is applied. All uncolored edges must belong to a feature loop. When such edges are detected, a search is initiated from an unmarked edge, and the edge is subsequently marked. This continues until all feature edges are marked, completing the grouping and forming the feature lines. Particles are then uniformly distributed along the feature lines, with their size and number determined by the total length of the feature lines and the specified particle spacing. If a feature line resembles a polyline, it can logically be treated as a straight line to ensure uniform particle generation. Typically, the size of the particle spacing does not perfectly divide the length of a feature line. Therefore, the particle spacing along the feature line will be slightly smaller to ensure a uniform distribution of particles.

\subsubsection{Multi-geometry Boolean operations} \label{Multi-geometry Boolean operations}

Boolean operations help define complex geometric boundaries and regions more precisely, especially for scenarios where interactions between different objects need to be simulated. Existing methods primarily rely on pre-processing software for boolean operations on the STL, requiring careful adjustment and optimization. In contrast, FPPG supports automatic geometric particle Boolean operations without prior STL processing.

\begin{figure}[pos=H]
\centering
\includegraphics[scale=0.7]{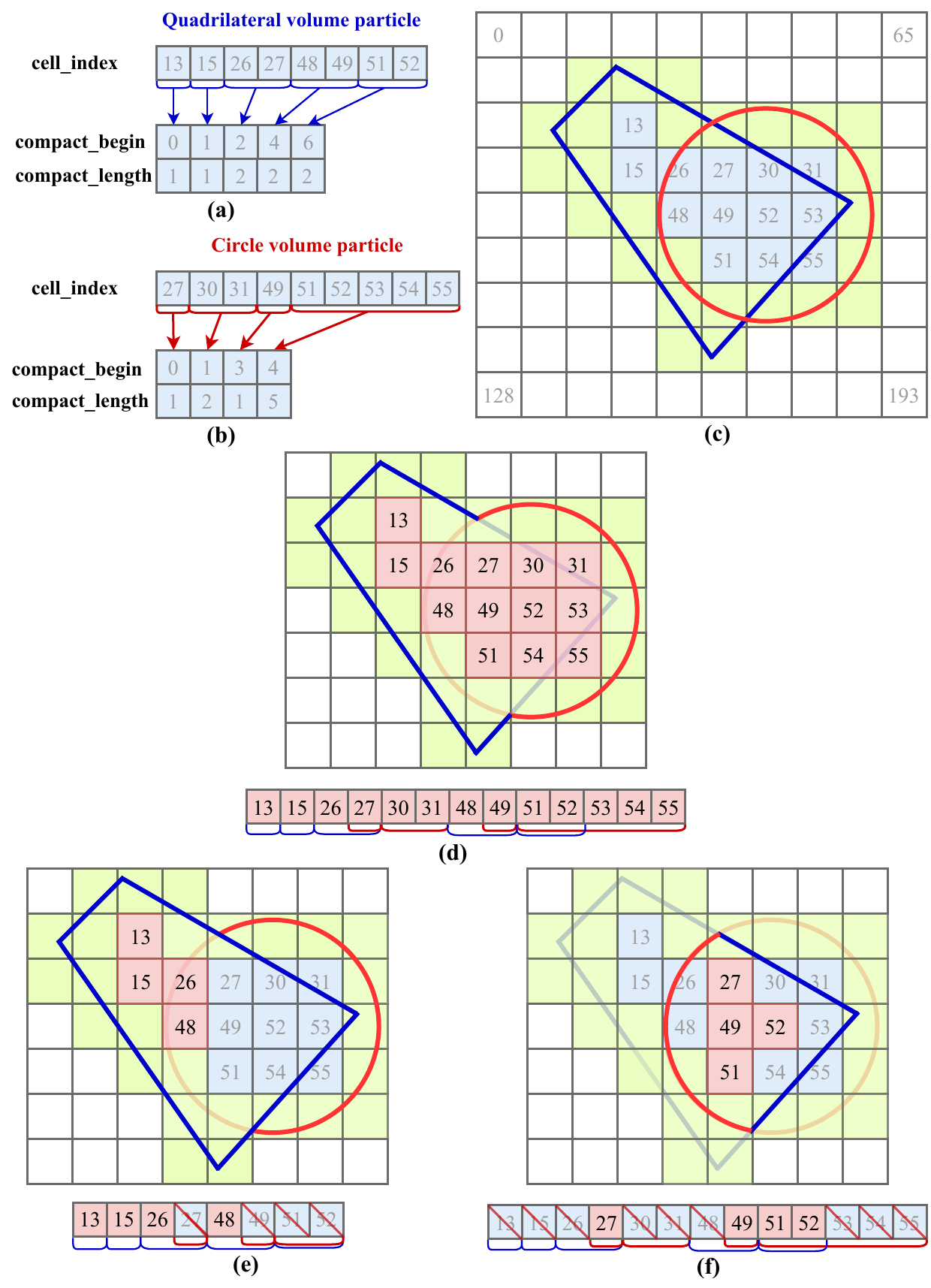}
\caption{Quadrilateral and Circle boolean operations. (a) Quadrilateral volume particle data structure. (b) Circle volume particle data structure. (c) Position relations and Morton encoding of partial grids. (d) Union operations. (e) Difference operation. (f) Intersection operation.}\label{fig-boolean}
\end{figure}

In FPPG, since surface and interior grids are defined according to the feature system outlined in Section \ref{feature system}, each geometry has its own set of surface grids and interior grids. Therefore, boolean operations between different geometries can be directly implemented using their respective grid sets straightforwardly. The $C_{compact}$ in the data structure stores the grid information corresponding to the geometry, and the boolean operations on the geometry are transformed into simple, etc. interval intersection, union, and grid value operations, e.g. Figure \ref{fig-boolean} shows a simple example with Boolean operations for quadrilateral and circle. Figure \ref{fig-boolean} a) and Figure \ref{fig-boolean} b) are the data structures for storing volume particles. The $C_{compact}^{begin}$ and $C_{compact}^{length}$ are regarded as the starting point and length of the interval. Results of Boolean operations on quadrilateral and circle volume particles can be obtained by simple union, intersection and difference of intervals.

\subsection{Physical relaxation for uniform particle optimization}

The initial particles consist of surface, internal and feature particles, which are not uniformly distributed at the current stage. In some regions, particle spacing is either too dense or too sparse to adequately represent the geometry, leading to increased simulation errors. To improve particle generation accuracy and computational efficiency, a physical-driven particle relaxation method is used for particle optimization. As described in Section 2, surface particles ($P_S$), internal particles ($P_I$), and feature particles ($P_F$) are defined, with different features that can exert forces on each other to prevent penetration during the relaxation process.

\subsubsection{Interparticle force} \label{interparticle force}
The evolution of particles following a physical relaxation process. In this study, it is necessary to solve the momentum equation for particles under a constant force field to achieve a uniform particle distribution. The governing equations of continuum mechanics in the Lagrangian reference frame include the equations of mass conservation and momentum conservation:
\begin{equation}
    \frac{d\rho }{dt} = -\rho \triangledown \cdotp \mathbf{v}, \label{eq11}
\end{equation}
\begin{equation}
    \frac{d\mathbf{v} }{dt} = -\mathbf{F}_p , \label{eq12}
\end{equation}
where $\rho$ denotes the density,$v$ denotes the velocity vector and $F_p$ denotes the pressure. Pressure represents the force created by the difference in pressure within the fluid. It corresponds to the gradient of the pressure field, directed from regions of high pressure to regions of low pressure.

The SPH discretization of the momentum equation is derived using Lagrangian mechanics. In this algorithm, particle motion is non-dissipative and does not take potential energy into account. Therefore, it can be obtained according to the control equation:
\begin{equation}
    \frac{dv }{dt} = -\displaystyle\sum_{j}m_j(\frac{p_i+p_j}{\rho_i\rho_j})\frac{\partial W(r_{ij},h_{ij})}{\partial r_{ij}}\mathbf{e}_{ij} ,\label{eq13}
\end{equation}
where the mass of each particle $m$ in this study is equal and set to 1. Other parameters are the same as in Equation \ref{eq6}.

As the particles relax to the target distribution, to eliminate the governing equations, an appropriate equation of state (EOS) must be established to relate pressure to density:
\begin{equation}
    p=P_0\left(\frac{\rho}{\rho_t}\right)^\gamma . \label{eq14}
\end{equation}
Once equilibrium is reached, the pressure becomes constant, resulting in a uniform particle distribution. 

Substitute EOS into Eq.(\ref{eq13}):
\begin{equation}
    \frac{dv}{dt}=-\sum_{j}{m_j\left(\frac{p_0}{p_{t,i}^2}+\frac{p_0}{p_{t,j}^2}\right)\frac{\partial W\left(r_{ij},h_{ij}\right)}{\partial r_{ij}}\mathbf{e}_{ij}} . \label{eq15}
\end{equation}

In the SPH method, the smooth kernel function determines the consistency and accuracy of the particle approximation. The smooth kernel function used in this study is the WendlandQuintic kernel (C2) \cite{wendland1995piecewise}.

\begin{equation}
    W\left(r_{ij},h\right)=\left\{\begin{matrix}\alpha_d\left(1-\frac{q}{2}\right)^4\left(2q+1\right)&0\le q\le2&\ \\0&q>2&\ \\\end{matrix}\right. ,\label{eq16}
\end{equation}
where $q$ denotes the model dimension, $\alpha_d$ is the normalisation factor.
\begin{equation}
    \alpha_d=\frac{7}{4\pi h^2},\;\;\;\;d=2 .\ \ \ \ \ \ \ \ \ \ \ \ \alpha_d=\frac{21}{6\pi h^3},\;\;\;\;d=3  . \label{eq17}
\end{equation}
By substituting the particle information into Equation (\ref{eq15}), the force exerted on the particle during relaxation can be calculated.

\subsubsection{Particle velocity and displacement} \label{particle velocity and displacement}

In conventional SPH simulations, the Velocity Verlet time integration algorithm is widely used\cite{ji2021feature}. However, in this study the simple velocity formulation in mechanics is sufficient. The time step is calculated using Equation (24) in \cite{ji2021feature}.
\begin{equation}
    \mathbf{v}_{n+1}=\mathbf{v}_n+\mathbf{a}_n\Delta t ,\label{eq22}
\end{equation}
where $\mathbf{a}_\mathbf{n}=\mathbf{F}_\mathbf{p}^\mathbf{n}$ is the acceleration and $\Delta t$ is the current time step. Note that particle evolution considers only the instantaneous acceleration. To achieve a completely stationary state, the particle velocity must be set to zero at the beginning of each time step.

Particle evolution under instantaneous acceleration is defined as:
\begin{equation}
    \mathbf{r}_{n+1}=\mathbf{r}_n+d\mathbf{r}=\mathbf{r}_n+\mathbf{v}_{n+1}\Delta t . \label{eq23}
\end{equation}

To maintain numerical stability and prevent particles from leaving the surface, the displacement is projected onto the surface of the geometric model:
\begin{equation}
    \mathbf{r}_{n+1}=\mathbf{r}_{n+1}-\frac{\mathbf{r}_{n+1}\cdot \mathbf{n}}{\left|\mathbf{n}\right|^2}\mathbf{n} .\label{eq24}
\end{equation}

Substituting Eq.(\ref{eq23}) into Eq. (\ref{eq24}) yields the displacement of the particle during relaxation:
\begin{equation}
    \mathbf{r}_{n+1}=\mathbf{r}_n+\mathbf{v}_{n+1}\Delta t-\frac{\left(\mathbf{r}_n+\mathbf{v}_{n+1}\Delta t\right)\cdot \mathbf{n}}{\left|\mathbf{n}\right|^2}\mathbf{n} , \label{eq25}
\end{equation}
where $\mathbf{n}$ is the vector normal to the triangular plane where the particle is located.

\subsubsection{Surface particle relaxation} \label{surface particle relaxation}
During the relaxation of surface particles, two types of forces are involved. The first is the internal interaction force between surface particles, which promotes uniform distribution. The second is the force exerted by feature line particles on surface particles, preventing them from leaving the surface.

Projecting displacements onto the surface effectively reduces the likelihood of particles moving away from the surface. However, this approach cannot entirely eliminate the issue, as surface particle migration may still occur. Therefore, it is necessary to inspect each particle to determine whether it remains on the surface. If not, it needs to be projected onto the surface again. This process is similar to the projection process described in Section \ref{mapping and body-fitted particles}. However, If projection fails, the particle should not be discarded to avoid particle loss. Instead, the nearest projectable adjacent triangle must be found again. Starting with the grid containing the particle, all adjacent grids (9 grids in 2D and 27 grids in 3D) are searched. The distances between the particle and the triangles in these grids are calculated and sorted. A check is performed to determine whether the particle can be projected onto the closest triangle. If none of the triangular faces can be projected, the particle has undergone excessive displacement and must return to its previous position.

The relaxation process of surface particles is illustrated in Algorithm \ref{alg1}. 

\begin{algorithm}
\caption{Surface particle relaxation}\label{alg1}
\begin{algorithmic}[1]
\Require Surface particles, volume particles, feature particles, optimise the number of iterations $num\_opt$;
\State $i \gets 0$
\While {$i < num\_opt$} do
\State Calculate the internal forces acting on the surface particles;(Eq. \ref{eq15})
\State Calculate the forces between surface particles and feature line particles;(Eq. \ref{eq15})
\State Calculate the time step;(Eq. 24 in \cite{ji2021feature})
\State Update speed;(Eq. \ref{eq22})
\State Update location;(Eq. \ref{eq25})
\State Surface particle projections, updating data structures, resetting forces;
\EndWhile
\State Outputs particle information according to the specified format. 
\end{algorithmic}
\end{algorithm}

\subsubsection{Volume particle relaxation} \label{volume particle relaxation}

\begin{figure}[pos=H]
\centering
\vspace{-5 pt}
\includegraphics[scale=.7]{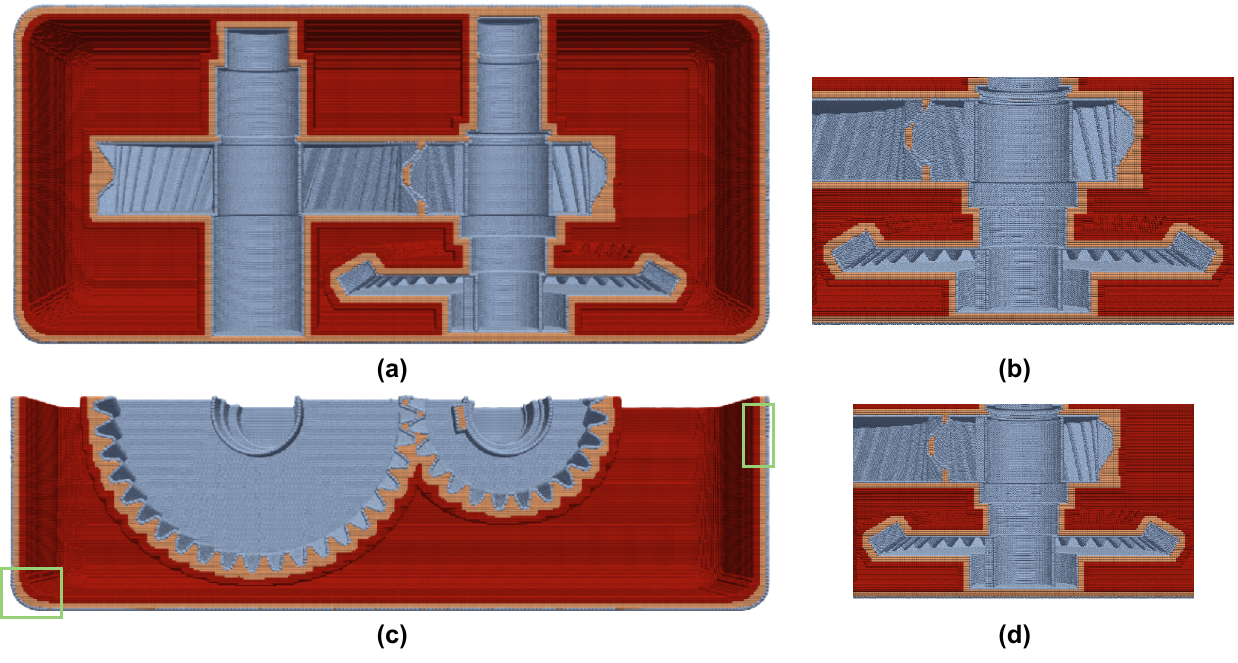}
\caption{Three particle layers are created in the gearbox volume. }\label{fig-volume}
\end{figure}

\begin{figure}[pos=H]
\centering
\includegraphics[scale=.75]{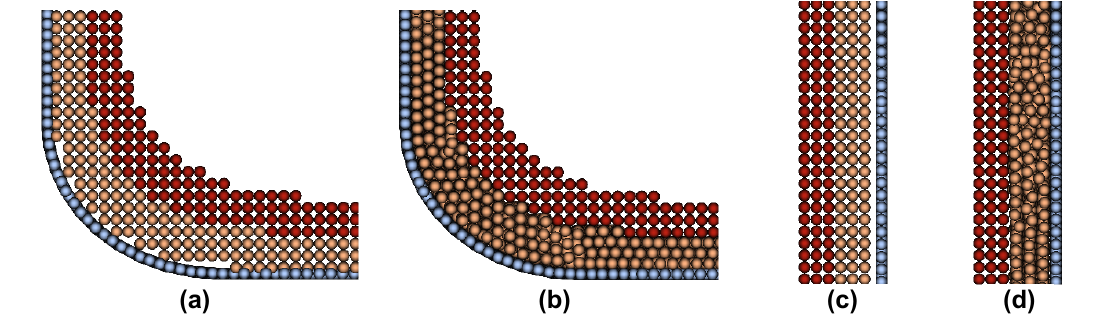}
\caption{A local comparison before and after optimization in Figure \ref{fig-volume} green box. (a) and (b) before optimization. (c) and (d) after optimization. }\label{fig-volume2}
\end{figure}

Volume particles near the surface must be uniformly distributed to minimize the risk of instability. In contrast, particles distant from the geometric surface maintain a stable core with an appropriate distribution and do not require relaxation optimization. Therefore, the relaxation process for volume particles focuses on the interior particles near the surface. Before volume particle relaxation is initiated, two particle layers must be pre-generated within the geometry, referred to as intermediate particles and innermost particles. Figure \ref{fig-volume} shows the three particle layers created for a gearbox, where the intermediate particles are the ones that need to be optimized. The width of each layer is user-defined. During relaxation, the intermediate particles are subjected to three types of forces: internal forces that promote target distribution and forces exerted by the surface and innermost particles to prevent leakage. Furthermore, the intermediate particles do not need to be mapped to the surface. Apart from force calculations and exclusion of mapping, the relaxation process for intermediate particles is similar to that of surface particles. All internal particle optimizations are also supported since the width is user-defined. The comparison before and after optimization is shown in Figure \ref{fig-volume2}. Before optimization, there is a gap between internal and surface particles. After optimization, the gap is closed and the particle distribution is more uniform.

\vspace{-20 pt}

\begin{figure}[pos=H]
\centering
\includegraphics[scale=.8]{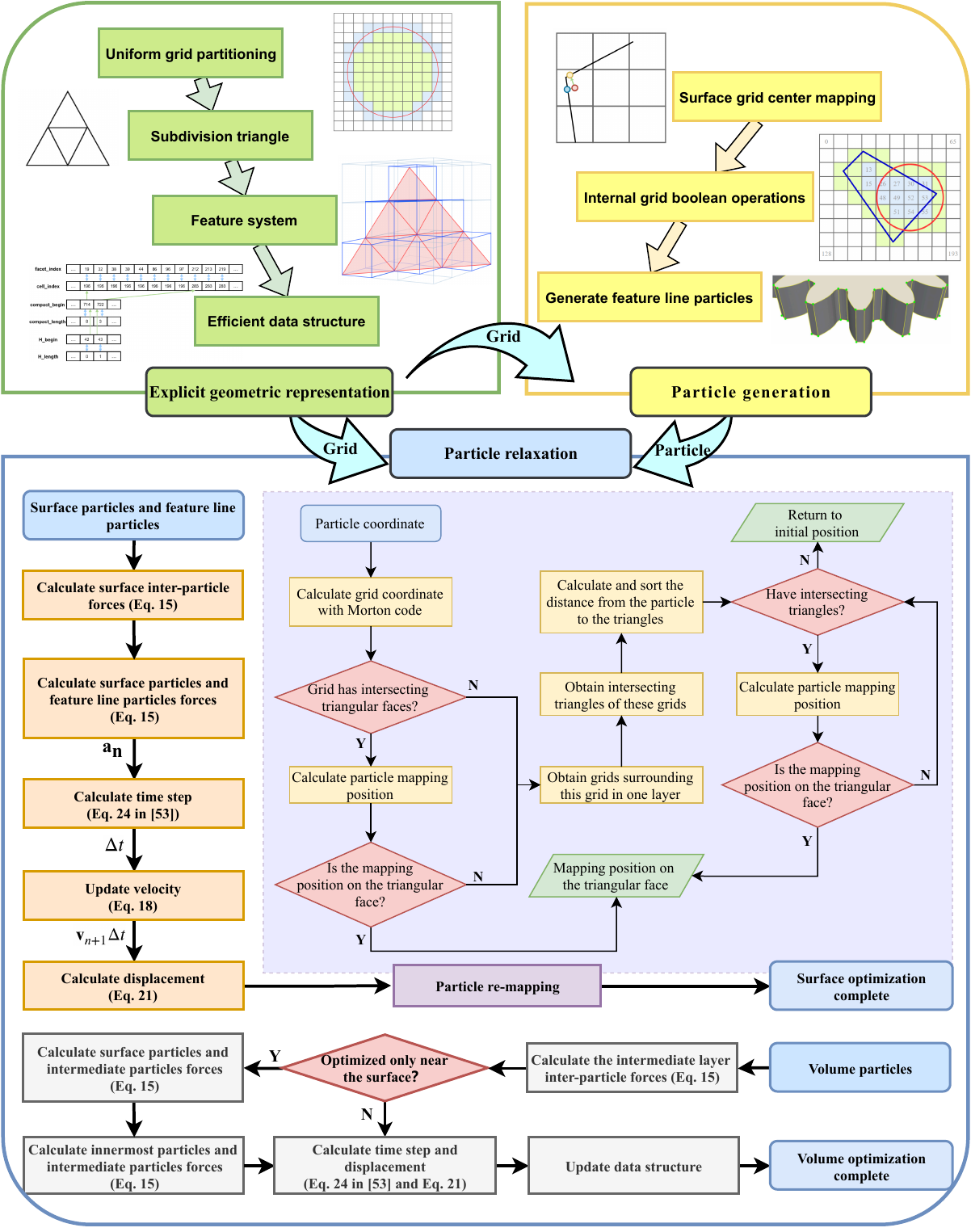}
\caption{The overall flow of the FPPG algorithm. }\label{fig-flow}
\end{figure}

\subsection{Algorithm review}

Figure \ref{fig-flow} provides a detailed description of the proposed method. Additionally, the Smoothed Particle Hydrodynamics (SPH) method inherently supports high parallelism as each grid is independent of others, thereby avoiding data conflicts and facilitating parallelization. OpenMP techniques are used to implement parallelization, and Chapter \ref{Numerical validations} elaborates on specific performance details.


\section{Numerical validations}\label{Numerical validations}
In this section, the proposed FPPG algorithm is validated using several numerical cases. The gear and tire models demonstrate the algorithm's ability to generate body-fitted and uniform particles on complex geometric surfaces while fully preserving features. The gearbox model confirms that the algorithm can handle multi-component structures by performing Boolean operations. In addition, FPPG is parallelized using OpenMP and performance tests are conducted on a multi-core shared memory architecture (AMD Ryzen 7 4800U with Radeon Graphics). An analysis is conducted on the efficiency of generation and optimization of particle sets of millions, tens of millions and even hundreds of millions, which exhibits the parallel scalability of the algorithms. A comparative analysis is also carried out using three common initial particle generation methods, highlighting the advantages of FPPG in terms of efficiency, shape-preservation ability and robustness, as shown in Table \ref{tab-compare}. Finally, applying the generated particles in industrial-scale computational fluid dynamics (CFD) simulations demonstrates practical utility in industrial applications. To facilitate reproducibility, the corresponding models and parameter settings are provided in each chapter.

\subsection{Complex geometries}
\begin{figure}[pos=b]
\centering
\includegraphics[scale=.73]{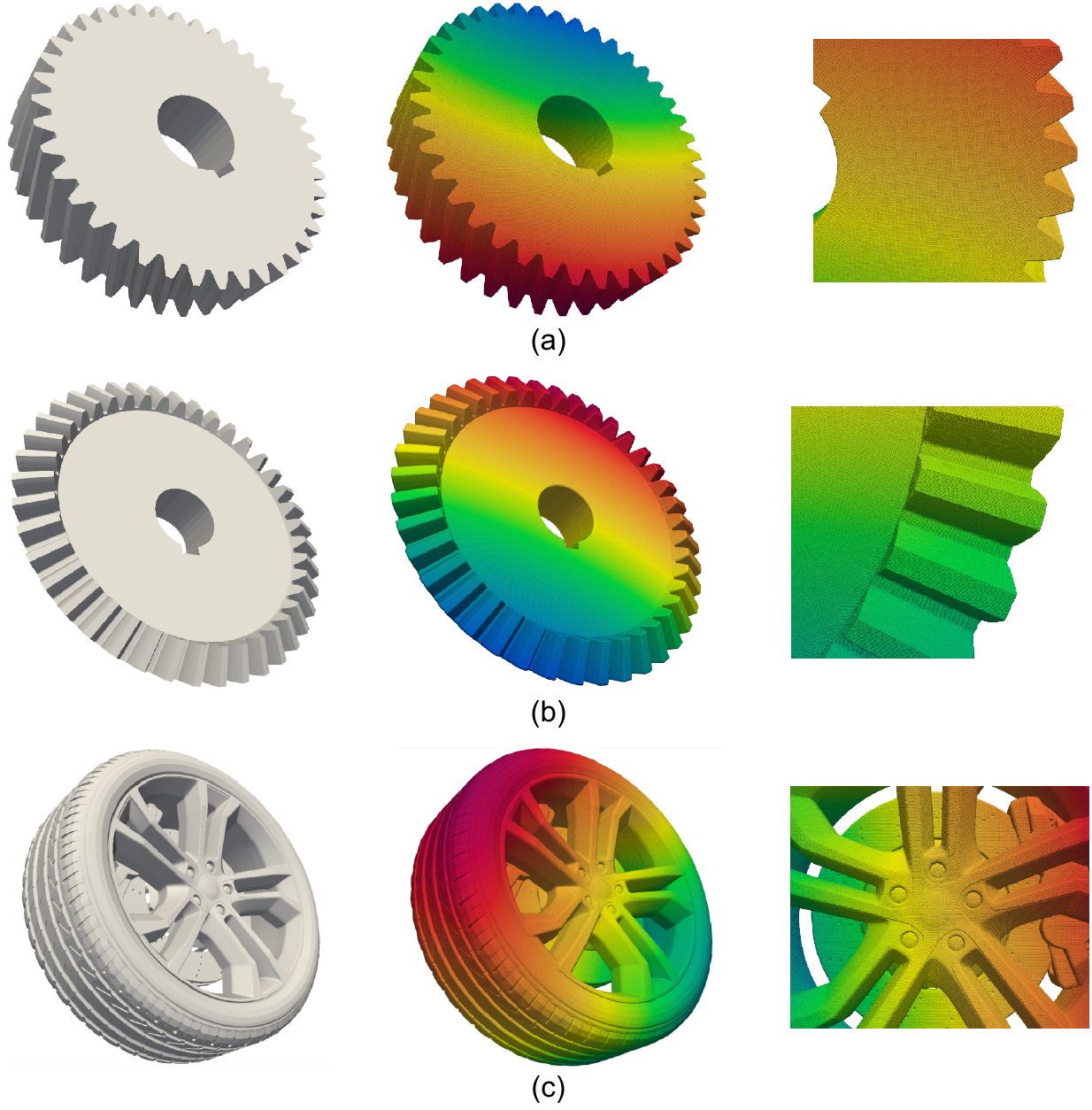}
\caption{Particle generation on geometric surfaces with fine features. (a) Gear 1. (b) Gear 2. (c) Tire. From left to right are CAD geometry in STL format, a global view of the generated particles, and a local view of the generated particles.}\label{fig-4-1}
\end{figure}

Particles were generated on the surfaces of three complex models, including two types of gears and a tire. These models contain sharp edges, singularities and other fine geometric details. Model parameters are listed in Table \ref{tab-complex geometry}. Figure \ref{fig-4-1} shows the original STL models on the left, the generated surface particles in the middle, and the enlarged details on the right, fully preserving the structural details. Notably, the particles generated for these gears and the tire will support a subsequent study to accurately simulate fluid flow over complex surfaces. These two industrial cases are discussed in detail in Section \ref{Industrial-level real case simulation}.

\begin{table}[width=.8\linewidth,cols=4,pos=h]
\renewcommand\arraystretch{1.5}
\caption{\label{tab-complex geometry} Three complex model parameters.}
\begin{tabular*}{\tblwidth}{@{\hspace{0.5cm}} CCCC @{\hspace{0.5cm} }}
\toprule
  Model & Model size & Particle diameter  & Particle number \\
\hline
Gear 1  & 134×30×134 & 0.3  & 682,049\\

Gear 2  & 143×29.9×143 & 0.15  & 2,236,266\\

Tire  & 0.72×0.25×0.72 & 0.002  & 866,573\\
\bottomrule
\end{tabular*}
\end{table}

\clearpage

\subsection{Boolean operations}
\begin{figure}[pos=H]
\centering
\includegraphics[scale=.25]{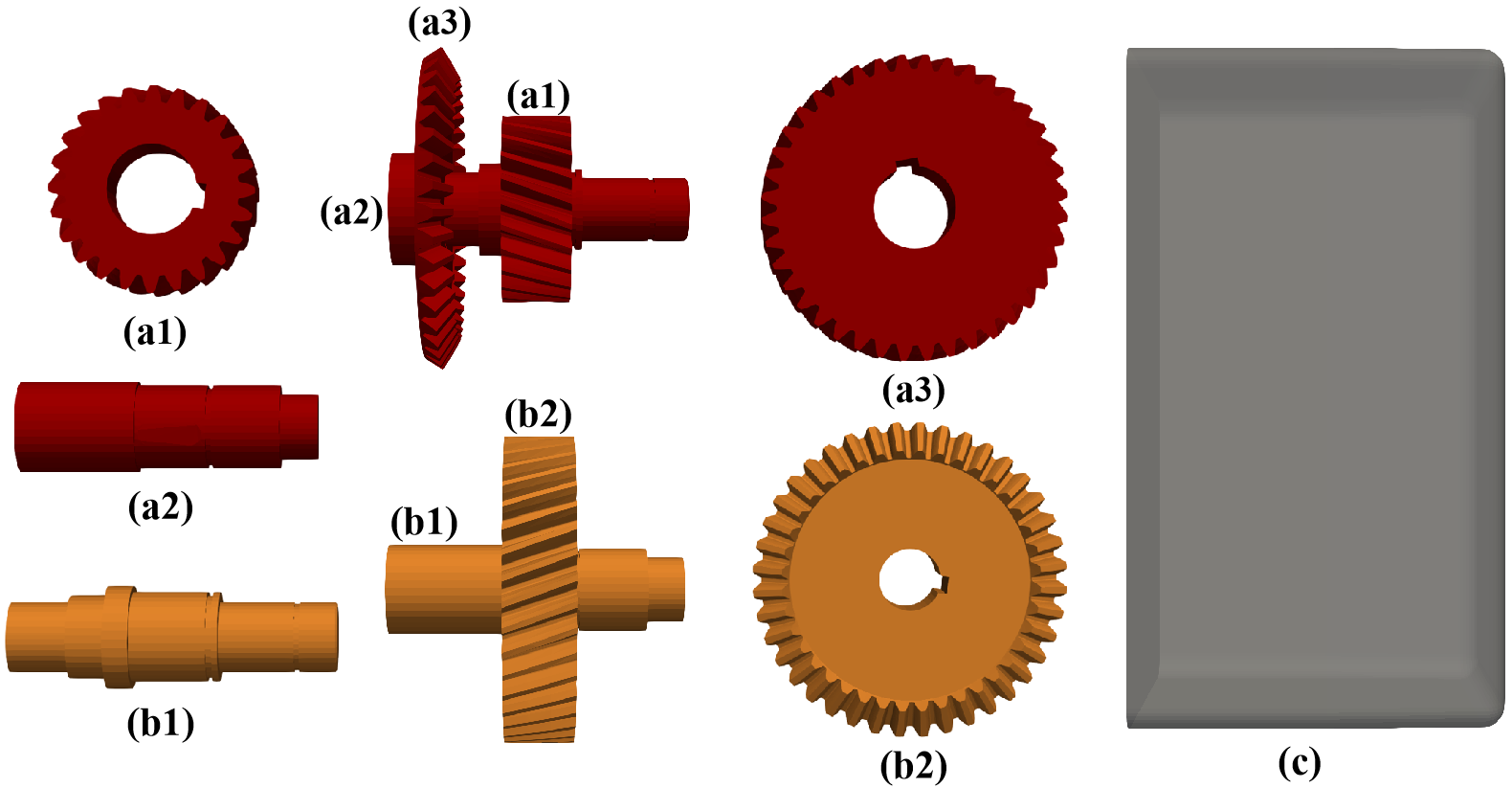}
\caption{Boolean operations model. (a1), (a2) and (a3) union operation. (b1) and (b2) union operation. (a1), (a2), (a3), (b1) and (b2) difference to (c).}\label{fig-boolean_test1}
\end{figure}

\begin{figure}[pos=H]
\centering
\includegraphics[scale=.23]{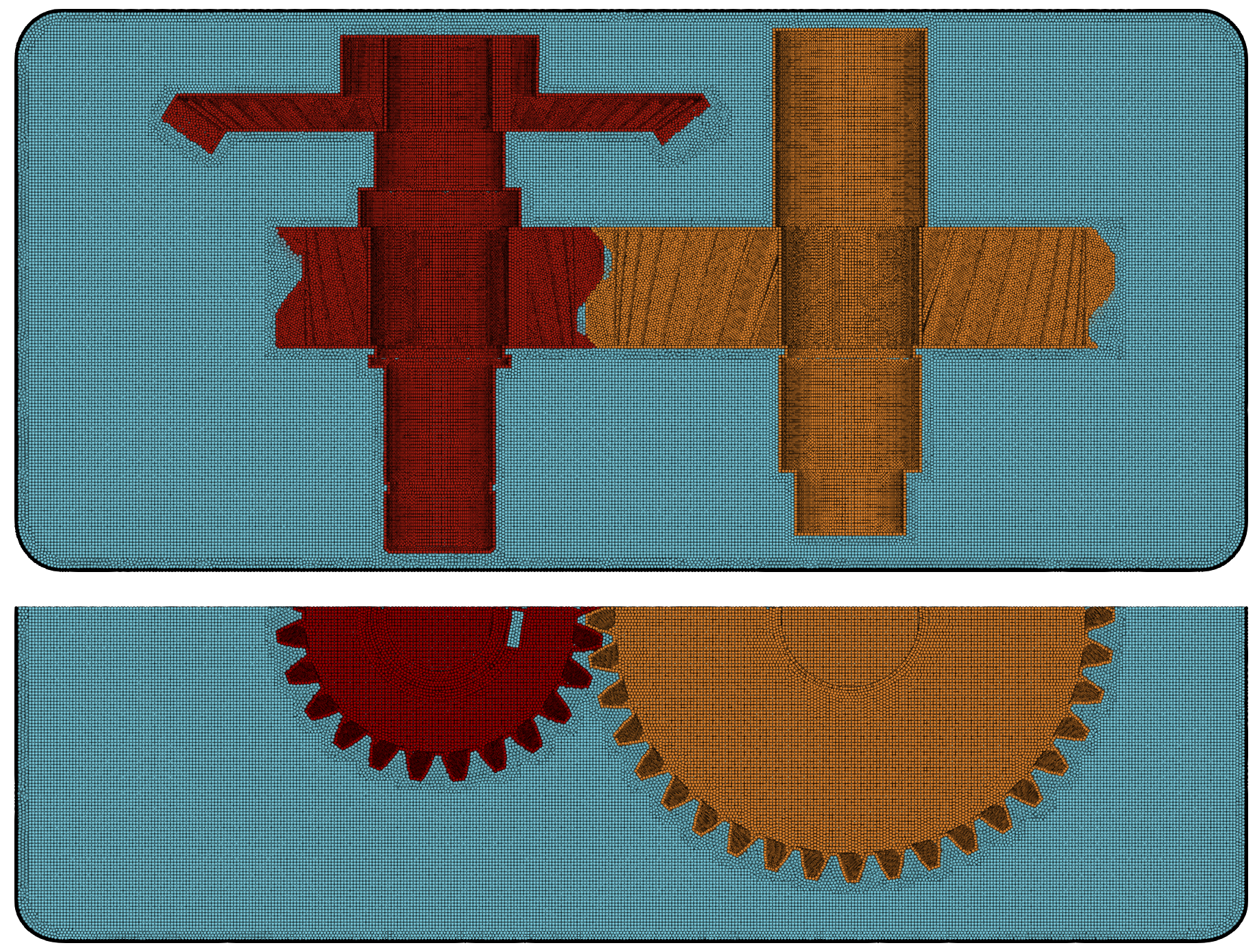}
\caption{Boolean operation result. }\label{fig-boolean_test2}
\end{figure}

Without preprocessing the model, FFPG can directly perform Boolean operations when generating particles. Figure \ref{fig-boolean_test1} illustrates a test model demonstrating a union operation applied to gears and shafts, as well as a difference operation between two sets of gears and a box. The results of the FPPG Boolean operation are shown in Figure \ref{fig-boolean_test2}.

\subsection{Parallelism and scalability}

\begin{table}
\small
\renewcommand\arraystretch{1.05}
\caption{\label{tab-generate} The speedup and efficiency.}
\begin{tabular*}{\tblwidth}{@{} CCCCCCCCCC@{} }
\toprule
\multirow{2}{*}{Geometry}        & \multirow{2}{*}{\makecell{Particle \\ diameter}}            & \multirow{2}{*}{\makecell{Particle \\ number}}                  & \multirow{2}{*}{\makecell{Thread \\ number}} &  \multicolumn{3}{c}{Particle generation}  & \multicolumn{3}{c}{Particle optimization (once)} \\
      &            &                   &  & Time(s) & Speedup & Efficiency & Time(s) & Speedup & Efficiency \\
\hline
\multirow{18}{*}{\makecell{Sphere \\ (1×1×1)}} & \multicolumn{1}{c}{\multirow{6}{*}{0.003}} & \multirow{6}{*}{1,287,443} & 1 & 4.05183 & - & - & 3.76095 & - & -  \\
                   & \multicolumn{1}{c}{}                   &                    & 2 & 2.38413 & 1.69950 & 84.98\% & 2.27183 & 1.65547 & 82.77\% \\
                   & \multicolumn{1}{c}{}                   &                    & 4 & 1.31731 & 3.07584 & 76.90\% & 1.37411 & 2.73700 & 68.43\% \\
                   & \multicolumn{1}{c}{}                   &                    & 8 & 0.95909 & 4.22467 & 52.81\% & 0.74955 & 5.01759 & 62.72\% \\
                   & \multicolumn{1}{c}{}                   &                    & 16& 0.66788 & 6.06675 & 37.92\% & 0.45888 & 8.19591 & 51.22\% \\
                   & \multicolumn{1}{c}{}                   &                    & 32& 0.56322 & 7.19408 & 22.48\% & 0.37665 & 9.98523 & 31.20\% \\
\cline{2-10}
  & \multicolumn{1}{c}{\multirow{6}{*}{0.001}} & \multirow{6}{*}{11,615,585} & 1 & 34.31428 & - & - & 38.34905 & - & - \\
                   & \multicolumn{1}{c}{}                   &                    & 2& 19.92611 & 1.72208 & 86.10\% & 22.09358 & 1.73576 & 86.79\% \\
                   & \multicolumn{1}{c}{}                   &                    & 4& 10.78055 & 3.18298 & 79.57\% & 12.59667 & 3.04438 & 76.11\% \\
                   & \multicolumn{1}{c}{}                   &                    & 8 & 9.85106 & 3.48331 & 43.54\% & 7.869665 & 4.87302 & 60.91\% \\
                   & \multicolumn{1}{c}{}                   &                    & 16& 5.66451 & 6.05777 & 37.86\% & 5.177340 & 7.40710 & 46.29\% \\
                   & \multicolumn{1}{c}{}                   &                    & 32& 4.55592 & 7.53180 & 23.54\% & 3.956014 & 9.69386 & 30.29\% \\
\cline{2-10}
  & \multicolumn{1}{c}{\multirow{6}{*}{0.0003}} & \multirow{6}{*}{129,219,285} & 1 & 413.39355 & - & - & 422.72243 & - & - \\
                   & \multicolumn{1}{c}{}                   &                    &2& 230.68081 & 1.79206 & 89.60\% & 244.61597 & 1.72811 & 86.41\% \\
                   & \multicolumn{1}{c}{}                   &                    &4& 121.17214 & 3.41162 & 85.29\% & 124.65559 & 3.39112 & 84.78\% \\
                   & \multicolumn{1}{c}{}                   &                    & 8& 92.11404 & 4.48785 & 56.10\% & 92.71626 & 4.55931 & 56.99\% \\
                   & \multicolumn{1}{c}{}                   &                    &16& 62.28473 & 6.63716 & 41.48\% & 55.72672 & 7.58563 & 47.41\% \\
                   & \multicolumn{1}{c}{}                   &                    &32& 43.43289 & 9.51798 & 29.74\% & 34.96091 & 12.09129 & 37.79\% \\
\cline{1-10}
\multirow{18}{*}{\makecell{Stanford Bunny  \\ (108×87×107)}} & \multicolumn{1}{c}{\multirow{6}{*}{0.1}} & \multirow{6}{*}{2,566,278} & 1 & 9.78440 & - & - & 11.47668 & - & - \\
                   & \multicolumn{1}{c}{}                   &                    & 2 & 5.77567 & 1.69407 & 84.70\% & 5.94139 & 1.93165 & 96.58\% \\
                   & \multicolumn{1}{c}{}                   &                    & 4 & 3.21914 & 3.03944 & 75.99\% & 3.06911 & 3.73941 & 93.49\% \\
                   & \multicolumn{1}{c}{}                   &                    & 8 & 2.35970 & 4.14645 & 51.83\% & 1.96965 & 5.82674 & 72.83\% \\
                   & \multicolumn{1}{c}{}                   &                    & 16& 1.55232 & 6.30308 & 39.39\% & 1.24532 & 9.21587 & 57.60\% \\
                   & \multicolumn{1}{c}{}                   &                    & 32& 1.42611 & 6.86090 & 21.44\% & 0.86952 & 13.1988 & 41.25\% \\
\cline{2-10}
  & \multicolumn{1}{c}{\multirow{6}{*}{0.03}} & \multirow{6}{*}{28,786,914} & 1 & 99.16436 & - & - & 135.66459 & - & - \\
                   & \multicolumn{1}{c}{}                   &                    & 2& 54.16624 & 1.83074 & 91.54\% & 69.91895 & 1.94031 & 97.02\% \\
                   & \multicolumn{1}{c}{}                   &                    & 4& 32.58243 & 3.04349 & 76.09\% & 37.36362 & 3.63093 & 90.77\% \\
                   & \multicolumn{1}{c}{}                   &                    & 8 & 28.73607 & 3.45087 & 43.14\% & 24.26470 & 5.59103 & 69.89\% \\
                   & \multicolumn{1}{c}{}                   &                    & 16& 16.91882 & 5.86119 & 36.63\% & 18.33339 & 7.39986 & 46.25\% \\
                   & \multicolumn{1}{c}{}                   &                    & 32& 13.01721 & 7.61794 & 23.81\% & 12.30709 & 11.02329 & 34.45\% \\
\cline{2-10}
  & \multicolumn{1}{c}{\multirow{6}{*}{0.01}} & \multirow{6}{*}{259,782,145} & 1 & 838.63147 & - & - & 1272.84190 & - & - \\
                   & \multicolumn{1}{c}{}                   &                    &2& 536.55547  & 1.56299 & 78.15\% & 673.23285  & 1.89064 & 94.53\% \\
                   & \multicolumn{1}{c}{}                   &                    &4& 293.15666  & 2.86069 & 71.52\% & 351.76674  & 3.61843 & 90.46\% \\
                   & \multicolumn{1}{c}{}                   &                    & 8& 193.90782 & 4.32490 & 54.06\% & 216.46313 & 5.88018 & 73.50\% \\
                   & \multicolumn{1}{c}{}                   &                    &16& 129.46119 & 6.47786 & 40.49\% & 125.40308 & 10.15000 & 63.44\% \\
                   & \multicolumn{1}{c}{}                   &                    &32& 93.47341  & 8.97187 & 28.04\% & 74.92988  & 16.98711 & 53.08\% \\
\bottomrule
\end{tabular*}
\end{table}

The FPPG method comprises two stages: particle generation and optimization. Particle generation occurs only once, while particle relaxation requires multiple iterations based on user-defined parameters. As the number of particles increases, the required optimization iterations also increase. In this section, different resolutions are applied to the sphere and Stanford Bunny models, generating and optimizing millions to billions of particles to evaluate the parallelism and scalability of the method. The model parameters and computation time statistics are presented in Table \ref{tab-generate}.

In the first stage of particle generation, speedup, and efficiency are listed in Table \ref{tab-generate}. Regardless of whether the particle number is in the millions, tens of millions, or billions, the FPPG method maintains a parallel efficiency exceeding $80\%$, demonstrating good parallel performance. Analyzing the parallel data for the sphere model, when the particle number reaches the million scale, two-thread efficiency is $84.98\%$. On the billion scale, the four-thread efficiency is $85.29\%$. As particle number and thread count increase, the parallel efficiency remains stable. Although the Stanford Bunny model is more irregular than the sphere model, it requires more computational resources for particle generation. However, the efficiency improves with increasing computational scale, indicating the scalability of the FPPG method during particle generation.

In the second stage of particle optimization, speedup, and efficiency are listed in Table \ref{tab-generate}. For the sphere model, parallel efficiency exceeds $80\%$, and for the Stanford Bunny model, it reaches over $90\%$. Similarly, in the parallel data for the sphere model, when the particle number reaches the million scale, two-thread efficiency is $82.77\%$. At the billion scale, four-thread efficiency reaches $84.78\%$. As particle number and thread count increase, the parallel efficiency remains stable. In the Stanford Bunny model, efficiency across multiple processing units improves with increasing computational scale, indicating that the FPPG method exhibits strong parallelism and scalability in particle relaxation optimization.

\subsection{Comparison with the feature-aware SPH method}

\begin{figure}[pos=H]
\centering
\includegraphics[scale=.3]{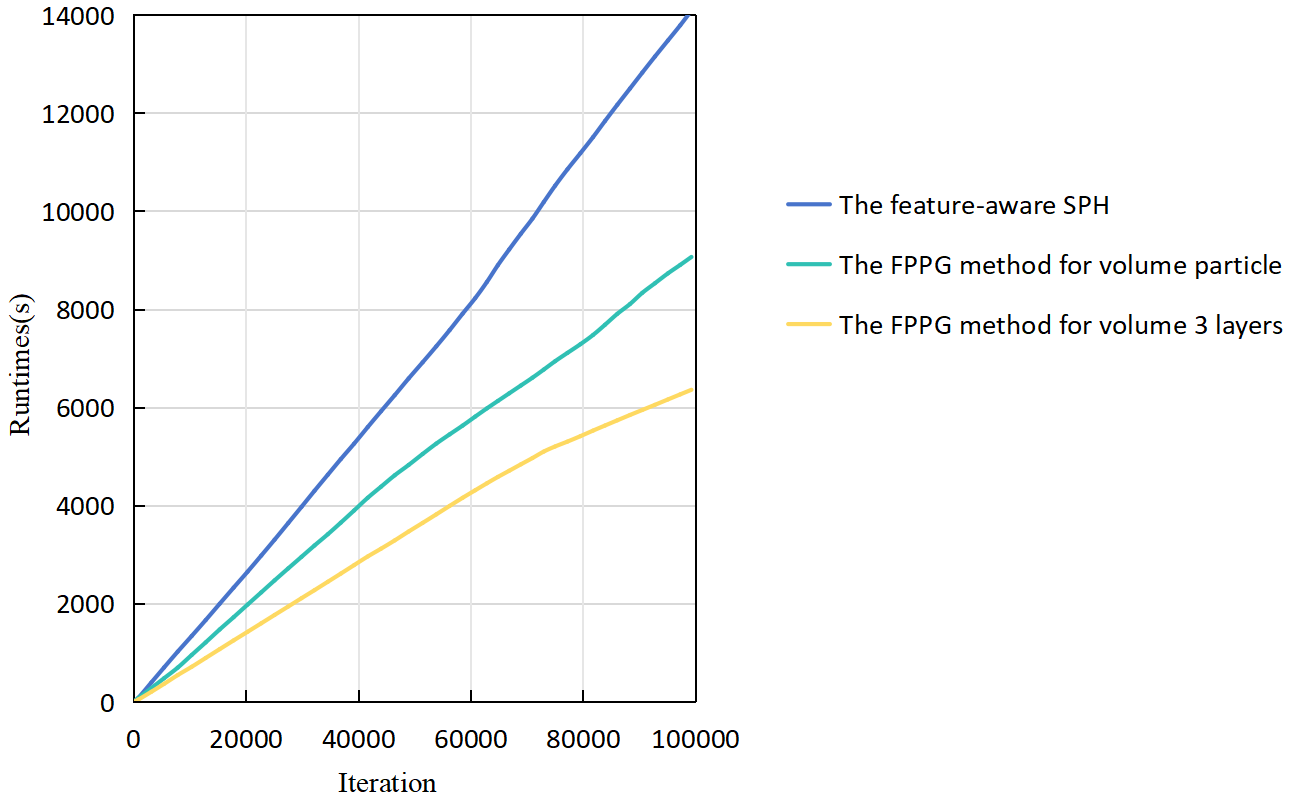}
\caption{Iteration time statistics.}\label{fig-feature-aware method2}
\end{figure}

Figure \ref{fig-feature-aware method2} presents the computation time for varying iteration counts using the feature-aware SPH and FPPG methods. The data for the feature-aware SPH method comes from Figure 9 of the reference \cite{ji2020consistent}, with a total of 240,370 particles. The FPPG method supports both full optimization of volume particles and only particles near the surface. Similarly, particles were generated on the Stanford Bunny model with a total of 240,692 particles (surface: 35,531; volume: 205,161). With the same number of iterations, the FPPG method consistently requires less computing time than the feature-aware SPH method. As the number of iterations increases, the time difference becomes more pronounced. Specifically, the feature-aware SPH method takes 14,097 seconds with 992,221 iterations, while the volume particle FPPG method takes 9,071 seconds and the three-layer FPPG method takes 6,362 seconds.		

\begin{figure}
\centering
\includegraphics[scale=.75]{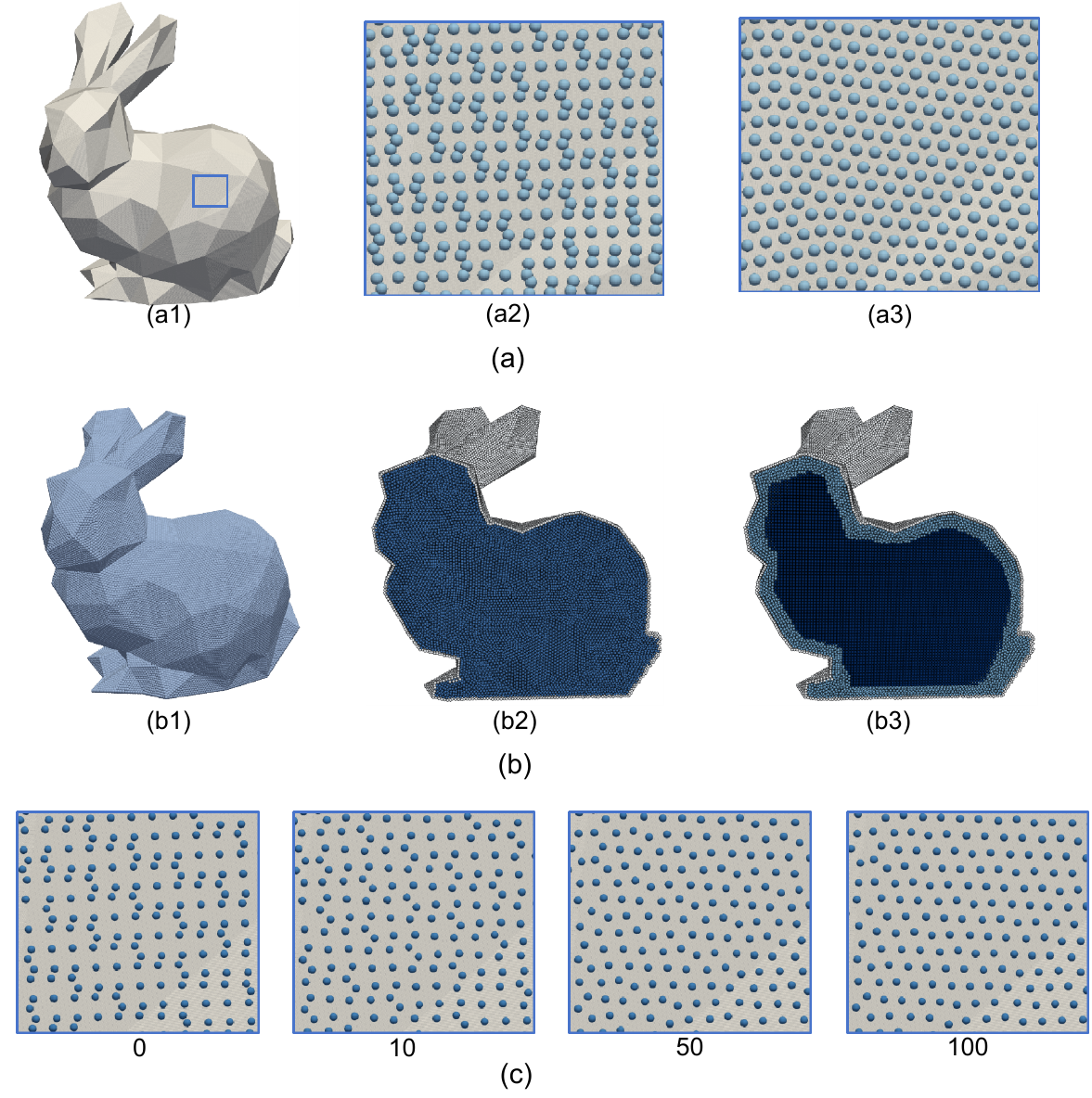}
\caption{FPPG generates particles at Stanford Bunny. (a) Surface particles. (b)Volume particles. (c) Different iterations of surface particles. }\label{fig-feature-aware method1}
\end{figure}

Figure \ref{fig-feature-aware method1} illustrates the particle generation on the Stanford Bunny model using the FPPG method. Figure \ref{fig-feature-aware method1} a) shows the surface particles generated on the Stanford Bunny, where Figure \ref{fig-feature-aware method1} a2) represents the initial surface particles and Figure \ref{fig-feature-aware method1} a3) shows the surface particles after 100,000 iterations, with the particle distribution becoming more uniform. Figure \ref{fig-feature-aware method1} b) demonstrates the volume particles generated within the Stanford Bunny model, where Figure \ref{fig-feature-aware method1} b2) shows the full optimization and Figure \ref{fig-feature-aware method1} b3) shows the optimization of spatial particles near the surface, both resulting in uniform distribution. Notably, the FPPG method achieves uniform distribution without requiring excessive iterations. Figure \ref{fig-feature-aware method1} c) presents magnified views at 0, 10, 50 and 100 iterations, with the distribution becoming relatively uniform after 100 iterations.

\subsection{Comparison with level-set based pre-processing techniques for particle methods}

\begin{figure}[pos=H]
\centering
\includegraphics[scale=.7]{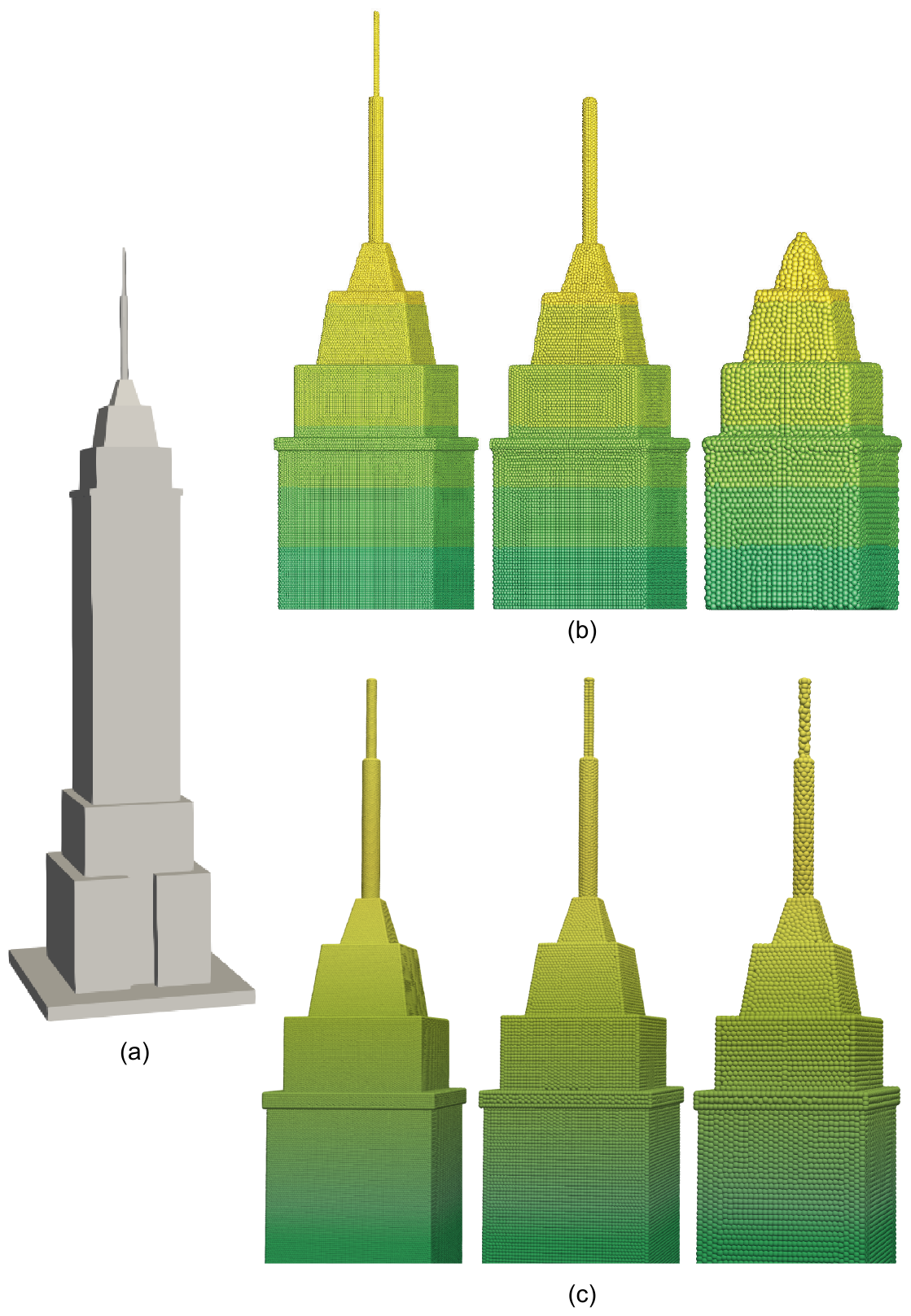}
\caption{(a) Skyscraper model. (b) Particles are generated using the Yu et al. proposed method. (c) Particles are generated using the FPPG method. }\label{fig-level-set}
\end{figure}

Yu et al. \cite{yu2022level} proposed a level-set-based preprocessing technique to identify and remove small fragments of "dirty" geometry that distort the shape at a given resolution. This method ensures the generation of body-fitted and uniform particle distributions for complex geometries. In Figure \ref{fig-level-set} a), particles are generated on a skyscraper model (6×6×21.6), with Figure \ref{fig-level-set} b) sourced from \cite{yu2022level}, and Figure \ref{fig-level-set} c) showing particles generated using the FPPG method. From left to right the resolutions are 0.03, 0.05 and 0.08. As resolution increases, the geometric details at the top of the model in Figure \ref{fig-level-set} b) are cleared to generate coarser resolution particles. In contrast, in Figure \ref{fig-level-set} c), the particles are generated correctly, with the geometry completely preserved.


Table \ref{tab-level-set} presents the computation time for 1000 iterations at varying resolutions. The FPPG method generates 5,879,300 particles in less than a second, whereas the method proposed by Yu et al. takes 146.98 seconds to generates 4,873,000 particles. For 1000 iterations, the FPPG method completes the process in only 358.47 seconds, significantly faster than the method of Yu et al. The FPPG method avoids particle deletion, ensuring a higher particle count while increasing computational efficiency.

\begin{table}[width=.9\linewidth,cols=5,pos=h]
\renewcommand\arraystretch{1.5}
\caption{\label{tab-level-set} The time required to generate particles with varying resolutions using Yu et al. method and FPPG method on the skyscraper model.}
\begin{tabular*}{\tblwidth}{@{\hspace{0.5cm}} CCCCC @{\hspace{0.5cm}} }
\toprule
Particle diameter & Method & Particle number  & Generation time(s) & Iteration time(s)\\
\hline
 \multirow{2}{*}{0.08}  & Yu et al. proposed method & 271,000	  & 13.20  &  145.98 \\
\cline{2-5}
 & FPPG 	 & 320,413  & 0.13  & 68.31\\
\hline
\multirow{2}{*}{0.05}  & Yu et al. proposed method & 1,559,000	  & 54.61   &  900.38 \\
 \cline{2-5}
 & FPPG 	 & 1,264,528  & 0.30	  & 134.74\\
 \hline
 \multirow{2}{*}{0.03}  & Yu et al. proposed method & 4,873,000	  & 146.98   &  2874.13 \\
 \cline{2-5}
  & FPPG 	 & 5,879,300  & 0.86  & 358.47\\
 \bottomrule
\end{tabular*}
 \end{table}

\clearpage

\subsection{Comparison with GenCase}
GenCase is a lattice-based particle generator and is applied to the DualSPHysics particle generation module. DualSPHysics is one of the most established open-source SPH frameworks \cite{crespo2015dualsphysics}, developed and maintained by a project team spanning multiple universities across Europe and the Americas. Since its initial release in 2011, it has experienced continuous iterations and evolved into an advanced particle-based solver.

Table \ref{tab-gencase} compares particle generation for a complete vehicle model using GenCase and FPPG at resolutions of 1 mm, 2 mm and 5 mm. Both codes are executed under identical hardware conditions with 16 OpenMP threads. It is important to note that the computation times in the table do not include the time spent on iterative optimization. At a particle count of two million, the GenCase method is 3.2 seconds faster, but when the particle count reaches 20 million, the FPPG method is 5.5 seconds faster. At particle numbers on the order of 100 million, the GenCase method fails, while FPPG can still generate particles within 1 minute (53.2 s).

\begin{figure}[pos=H]
\centering
\includegraphics[scale=.75]{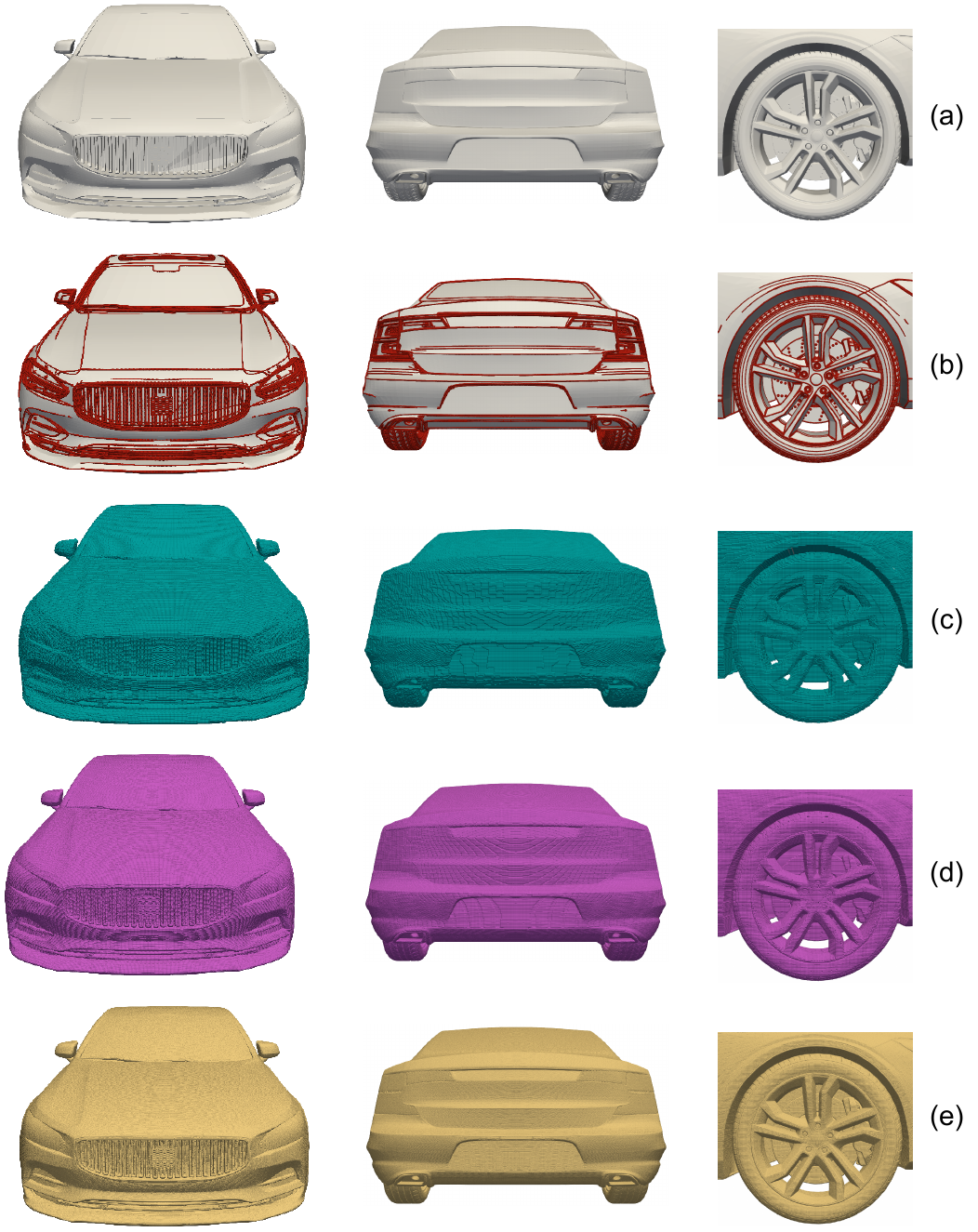}
\caption{From left to right, the front, back and tire of the vehicle are shown. (a) Vehicle model. (b)Feature particles. (c) Particles generated by GenCase. (d) Initial particles generated by the FPPG. (e) Optimized particles generated by the FPPG.}\label{fig-gencase}
\end{figure}

In addition, the most significant advantage of FPPG over GenCase is the quality of the generated particles. Figure \ref{fig-gencase} a) shows the vehicle model. Figure \ref{fig-gencase} b) shows FPPG-generated feature particles on the model surface. In contrast, Figure \ref{fig-gencase} c) shows the particles generated by GenCase, which exhibit a stair-stepping pattern on the surface with significant discretization errors and many inaccurately represented geometric features. Figure \ref{fig-gencase} d) shows the particles directly generated by FPPG before optimization, preserving more details of the original geometry. Finally, Figure \ref{fig-gencase} e) shows the FPPG optimized particles, which are uniformly distributed.

\begin{table}[width=1\linewidth,cols=4,pos=h]
\renewcommand\arraystretch{1.1}
\caption{\label{tab-gencase} Performance comparison of GenCase and FPPG in particle generation for vehicle models.}
\begin{tabular*}{\tblwidth}{@{} CCCC @{} }
\toprule
Particle diameter(mm) & Generator & Particle number  & computation time(s) \\
\hline
 \multirow{2}{*}{5}  & GenCase & 2,702,162  & 1.5 \\
 \cline{2-4}
 & FPPG & 2,720,184  & 4.7 \\
\hline
 \multirow{2}{*}{2}  & GenCase & 22,085,673  & 25.3 \\
 \cline{2-4}
 & FPPG & 22,773,444  & 19.8 \\
\hline
 \multirow{2}{*}{1}  & GenCase & Failure  & Failure \\
 \cline{2-4}
 & FPPGv & 101,485,542  & 53.2 \\
\bottomrule
\end{tabular*}
\end{table}

\begin{table}[width=1\linewidth,cols=3,pos=h]
\renewcommand\arraystretch{1.1}
\caption{\label{tab-compare} A comparative summary of FPPG with the feature-aware method, the method of Yu et al. and GenCase.}
\begin{tabular*}{\tblwidth}{@{} C  C  l  @{} } 
\toprule
Explicit method & Implicit method & \makecell[c]{FPPG advantage}   \\
\hline
 \multirow{7}{*}{FPPG}& \multirow{2}{*}{feature-aware method}  & \makecell[l]{The FPPG particle generation is quicker and has fewer iterations.}\\ 
 &&The overall computation time is therefore shorter. \\
 \cline{2-3}
 & \multirow{3}{*}{Yu et al. method}  & Larger resolution, the FPPG  preserves geometric features. \\
 &&The  Yu et al. method requires sacrificing some geometric details. \\
 &&For the same effect, FPPG uses fewer particles and achieves faster computation. \\
 \cline{2-3}
 & \multirow{2}{*}{GenCase}  &  The FPPG is more robust.\\ &&And the particles are shape-preserving and uniform. \\
\bottomrule
\end{tabular*}
\end{table}

\subsection{Industrial-level geometries and simulations} \label{Industrial-level real case simulation}
With the advancement of SPH, the industry has shown increasing interest in the SPH  method\cite{vacondio2021grand}. Engineers believe SPH is capable of handling applications with highly distorted and complex interfaces, such as gearbox and tire hydrodynamics, which are becoming more prevalent. The particles generated by FPPG can be used directly in computational fluid dynamics simulations with a simple and easily coupled interface. Vehicle wading and gearbox splash lubrication simulations are presented here to demonstrate the effectiveness of FPPG in improving particle distribution quality at the boundaries of complex geometric flow fields.

\subsubsection{Vehicle wading simulation}

Vehicle wading is a classic problem in automotive engineering. Traditional methods rely on physical testing using prototype vehicles that closely replicate the physical scenarios of wading but are often unable to capture the flow details of water entering the vehicle interior. Applying numerical simulations to study vehicle fording performance can help identify problems in the early stages of vehicle development.

\begin{figure}[pos=H]
\centering
\includegraphics[scale=.2]{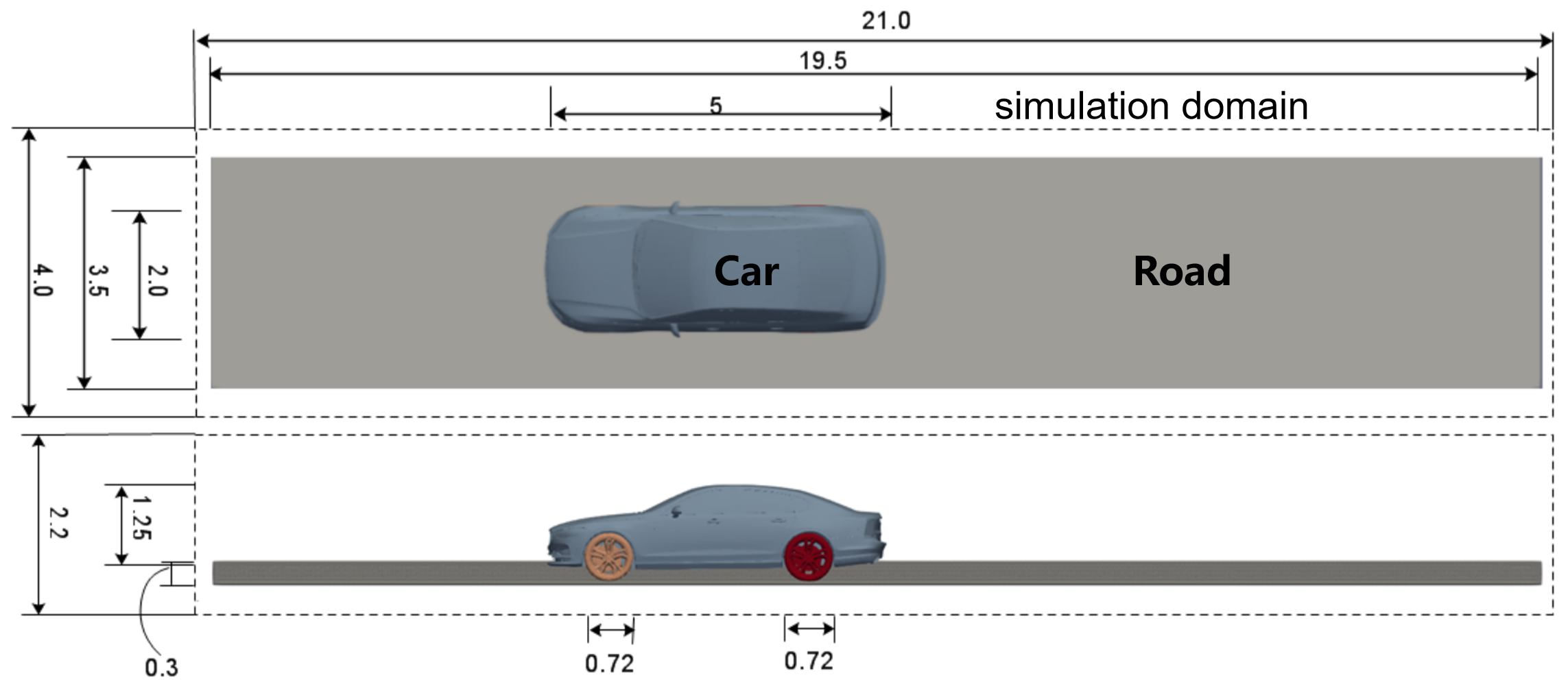}
\caption{Vehicle wading simulation model. }\label{fig14}
\end{figure}

\begin{figure}[pos=H]
\centering
\includegraphics[scale=.78]{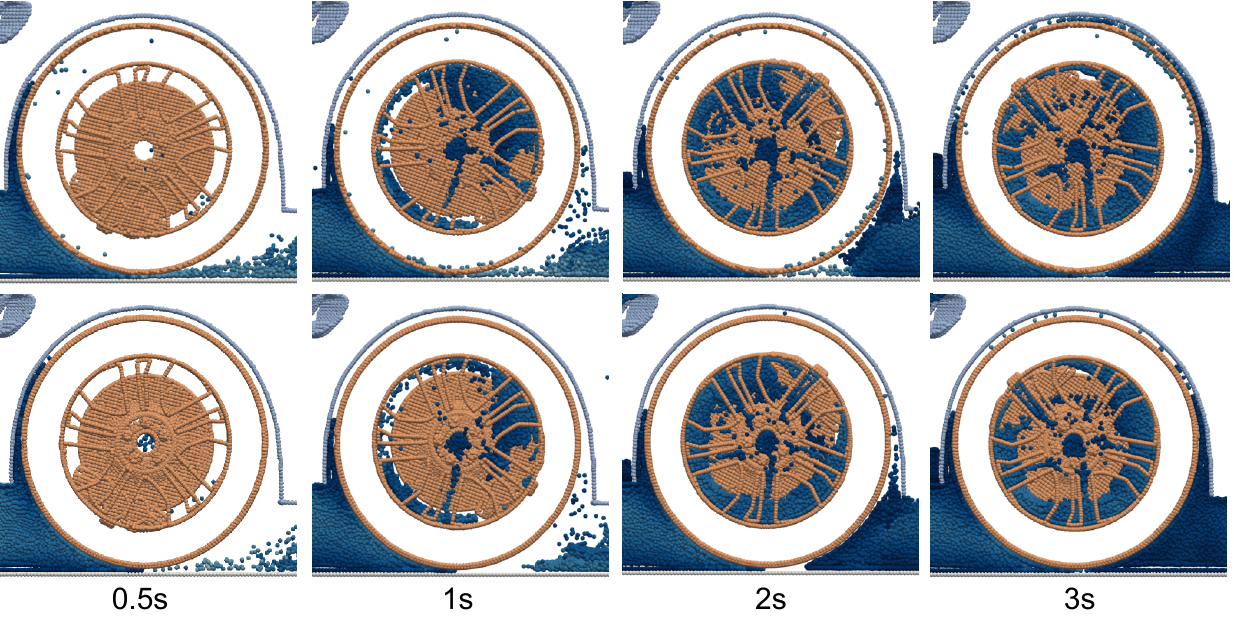}
\caption{The cross-sectional diagrams of wading simulation before and after particle optimization are shown from left to right, representing the simulation results at 0.5s, 1s, 2s and 3s, respectively. The upper part is not optimized and water enters the tire. The lower part is optimized, without infiltration}\label{fig17}
\end{figure}

Figure \ref{fig14} shows the automotive model alongside the simulated roadway, while Tables \ref{tab3} present the main parameter configurations during the simulation process. Figure \ref{fig17} presents cross-sections of the wheels at four different time intervals during the simulation. In the upper section, the surface particles of the model are not optimized, which leads to water particles penetrating the tire. Conversely, the lower section optimized the surface particles and eliminated the penetration phenomenon. The generated particles significantly mitigate the issue of water particle penetration in the simulation and successfully model the vehicle wading performance.

\begin{table}[width=.9\linewidth,cols=4,pos=h]
\caption{\label{tab3} Vehicle wading simulation parameters.}
\begin{tabular*}{\tblwidth}{@{} CCL@{} }
\toprule
Params & Value & Help  \\
\hline
gravity & [0,0,-9.81] & gravitational acceleration$(m/s^2)$ \\
speedsound & 100 & velocity$(m/s)$ \\
h & $0.5e^{-2}$ & smooth length$(m)$ \\
dt & $0.1e^{-3}$ & minimum computation time step$(s)$ \\
PresetVelocity & $5.0$ & initial velocity of water particles$(m/s)$ \\
AngularVelocity & $803.94$ & angular velocity of a wheel$(rad/s)$ \\
TimeOut & $0.05$ & output time interval$(s)$ \\
TimeMax & $5.0$ & simulation duration$(s)$ \\
\bottomrule
\end{tabular*}
\end{table}

\begin{table}[width=.9\linewidth,cols=4,pos=h]
\caption{\label{tab4} Gearbox splash lubrication simulation parameters.}
\begin{tabular*}{\tblwidth}{@{} CCL@{} }
\toprule
Params & Value & Help  \\
\hline
gravity & [0,0,-9.81] & gravitational acceleration$(m/s^2)$ \\
speedsound & 10 & velocity$(m/s)$ \\
h & $0.4e^{-3}$ & smooth length$(m)$ \\
dt & $0.1e^{-4}$ & minimum computation time step$(s)$ \\
PresetVelocity & $0$ & initial velocity of water particles$(m/s)$ \\
AngularVelocity & $3000$ & angular velocity of a wheel$(rad/s)$ \\
TimeOut & $0.008$ & output time interval$(s)$ \\
TimeMax & $0.8$ & simulation duration$(s)$ \\
\bottomrule
\end{tabular*}
\end{table}

\subsubsection{Gearbox splash lubrication simulation}

Traditional gearbox designs use transparent housing experiments to validate the splash lubrication effect and spatial distribution of lubricating oil. However, these experiments are costly, time-consuming, and challenging to measure the flow rate of lubricating oil in specific regions within the housing. Numerical simulations can significantly reduce the need for repeated manufacturing of transparent housing experiments.

\begin{figure}[pos=H]
\centering
\includegraphics[scale=.72]{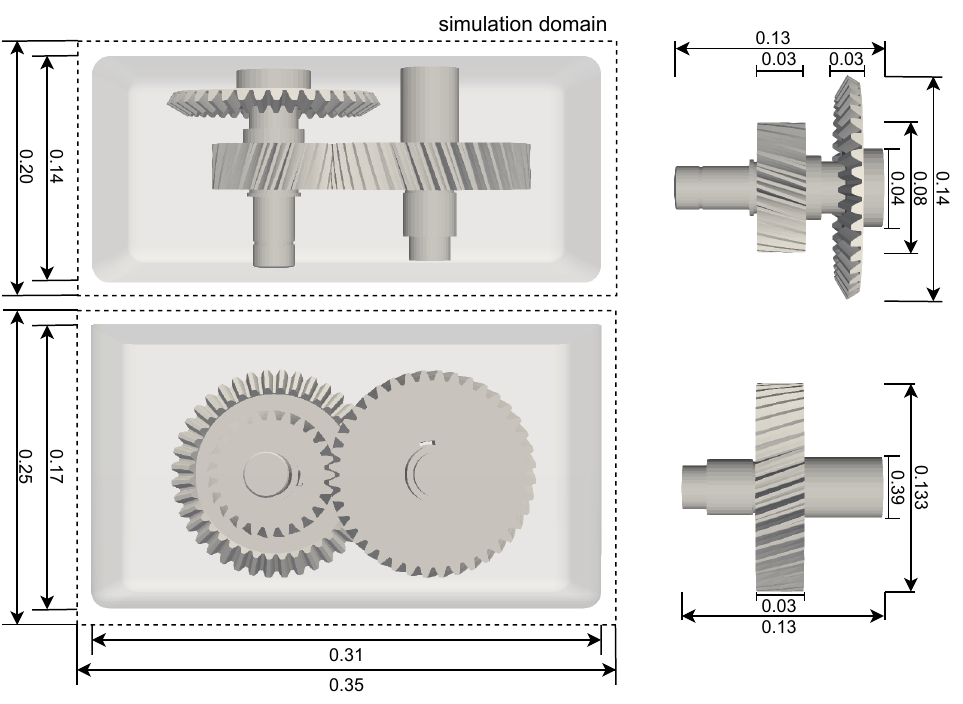}
\caption{Gearbox model. }\label{fig18}
\end{figure}

\begin{figure}[pos=H]
\centering
\includegraphics[scale=.6]{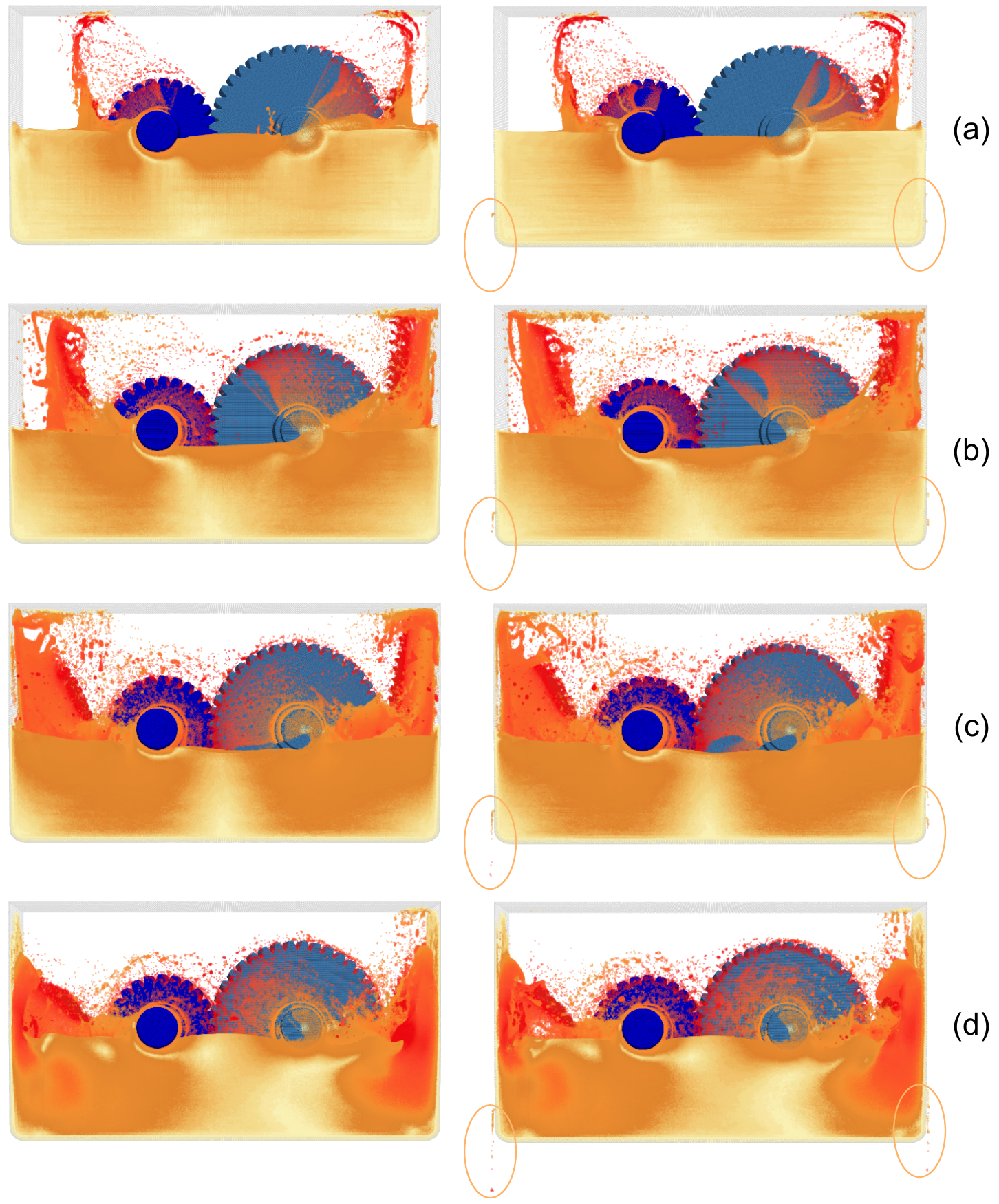}
\caption{(a) to (d) show the simulated cross-sections of the gearbox before and after optimization and represent the simulation results at 0.1 s, 0.4 s, 0.7 s and 1 s, respectively. Optimized particles are shown on the left and on the right unoptimized particles. Orange circles indicate particles leaking out of the gearbox. }\label{fig19}
\end{figure}

Figure \ref{fig18} illustrates the geometry, while Table \ref{tab4} presents the main parameter configurations used during the simulation process. Figure \ref{fig19} illustrates cross-sectional views of the gearbox splash lubrication simulation at four different time intervals. On the right, the model surface particles are not optimized, resulting in a non-uniform particle distribution that does not fit the geometry and causes fluid particles to leak out of the gearbox. In contrast, on the left, the optimized surface particles are uniformly distributed and fit the geometry, effectively maintaining shape and preventing particle overflow from the box. In this case, the splash lubrication behavior of the gearbox is successfully simulated.

\section{Conclusions}
In this paper, a feature-preserving particle generation (FPPG) method is proposed to achieve body-fitted and uniform particle generation based on explicit geometric representation. Various numerical validations are conducted to evaluate surface and volume particle generation for complex geometries with sharp features. The main contribution of the paper can be summarized as:
\begin{enumerate}
    \item FPPG employs explicit geometric representation for the generation and optimization of high-quality particles. Numerical tests of various complexity demonstrate that FPPG is able to  fully preserve geometric features, e.g. sharp edge, singularity point, and etc., without losing accuracy from the input geometry;
    \item FPPG is entirely parallelized using OpenMP to explore the computational capability of shared-memory architectures for rapid particle generation and optimization. Scalability tests show that good speedup and efficiency are achieved for both procedures, where a maximum speedup of 9.5X and 17.0X is achieved respectively;
\item FPPG is compared with other state-of-the-art algorithms. Comparison results suggest that FPPG is able to capture more detailed feature from the input geometry and requires remarkably less runtime in the meantime owing to its high-concurrency nature and the additional optimization techniques we introduced.
\item FPPG is validated further through two industrial cases. Results from the vehicle wading and gearbox splash lubrication case reveal that FPPG is capable and feasible for real engineering problems. With the accurate particle setup calculated by FPPG, the solver is able to deliver more reliable and trustworthy results.
\end{enumerate}

In terms of future development, we will further improve the performance of the proposed method by extending FPPG to GPU architectures. Additionally, extending the algorithm to support multi-resolution and address multi-scale complex problems is of significant interest too.

\section*{Acknowledgments}

Zhe Ji is supported by Guangdong Basic and Applied Basic Research Foundation (No. 2022A1515110314), the General Program of Taicang Basic Research Project (No.TC2022JC07), and the National Natural Science Foundation of China (Grant No.12301560).

\bibliographystyle{elsarticle-num-names}

\bibliography{cas-refs}

\end{document}